\documentclass{jfm}
\usepackage{graphicx}
\usepackage{epstopdf, epsfig}
\usepackage{comment}
\usepackage[figuresright]{rotating}
\usepackage{booktabs}

\usepackage{subcaption}
\usepackage{wrapfig,caption}
\usepackage{multirow}
\usepackage{amsmath}
\usepackage{upgreek}
\usepackage{hyperref}
\usepackage{fancyhdr}
\usepackage{enumitem}
\usepackage{color}
\usepackage{array}
\usepackage{amstext}
\usepackage{amsmath} 
\usepackage{amssymb}
\usepackage[usenames,dvipsnames,svgnames,table]{xcolor}
\usepackage{psfrag}
\usepackage{array}
\usepackage{threeparttable}
\usepackage{dcolumn}
\newcolumntype{d}{D{.}{.}{-1}}
\usepackage{enumitem}
\usepackage{titlesec}
\usepackage{float}
 \usepackage[numbers,sort,comma,square]{natbib}
 \usepackage{makecell}
\setcellgapes{4pt}%
\def\bm#1{\mbox{\boldmath{$#1$}}}

\newcommand\Reyn{\mbox{\textit{Re}}}  
\newcommand\Arc{\mbox{\textit{Ar}}}  
\newcommand\Fr{\mbox{\textit{Fr}}}  
\def\a{\hat{\mathbb{U}}_r}
\def\b{\hat{\mathbb{R}}_f}
\def\c{\hat{\mathbb{R}}_p}

\usepackage{xcolor}
\usepackage{tikz}
\usepackage{pgfplots}

\pgfplotsset{
        layers/my layer set/.define layer set={
            background,
            main,
            foreground
        }{
        },
        set layers=my layer set,
    }

\definecolor{findOptimalPartition}{HTML}{D7191C}
\definecolor{storeClusterComponent}{HTML}{FDAE61}
\definecolor{dbscan}{HTML}{ABDDA4}
\definecolor{constructCluster}{HTML}{2B83BA}

\definecolor{ColorOne}{HTML}{61BB46}
\definecolor{ColorTwo}{HTML}{FDB827}
\definecolor{ColorThree}{HTML}{F5821F}
\definecolor{ColorFour}{HTML}{E03A3E}
\definecolor{ColorFive}{HTML}{963D97}
\definecolor{ColorSix}{HTML}{009DDC}
\definecolor{Invisible}{HTML}{FFFFFF}

\usepackage{tikz}
\usepackage{blindtext}
\usetikzlibrary{plotmarks}

\usepackage{pgfplots}
\pgfplotsset{compat=newest}

\usetikzlibrary{plotmarks}

\newcommand{\LearnedCirc}{\raisebox{2pt}{\tikz{\filldraw[red](-.5,-1.50) circle(2pt);}}}
\newcommand{\LearnedSquare}{\raisebox{0pt}{\tikz{\filldraw[red](-0.15,-0.15) rectangle (1pt, 1pt);}}}

\newcommand{\TermOneBlock}{\raisebox{0pt}{\tikz{\filldraw[ColorOne](-0.25,-0.15) rectangle (1pt, 1pt);}}}
\newcommand{\TermTwoBlock}{\raisebox{0pt}{\tikz{\filldraw[ColorTwo](-0.25,-0.15) rectangle (1pt, 1pt);}}}
\newcommand{\TermThreeBlock}{\raisebox{0pt}{\tikz{\filldraw[ColorThree](-0.25,-0.15) rectangle (1pt, 1pt);}}}
\newcommand{\TermFourBlock}{\raisebox{0pt}{\tikz{\filldraw[ColorFour](-0.25,-0.15) rectangle (1pt, 1pt);}}}
\newcommand{\TermFiveBlock}{\raisebox{0pt}{\tikz{\filldraw[ColorFive](-0.25,-0.15) rectangle (1pt, 1pt);}}}
\newcommand{\TermSixBlock}{\raisebox{0pt}{\tikz{\filldraw[ColorSix](-0.25,-0.15) rectangle (1pt, 1pt);}}}

\newcommand{\TermOne}{\raisebox{-1pt}{\tikz{\draw[very thick, color=ColorOne, solid](0,0)--(6mm,0);\node[mark size=3pt,color=ColorOne] at (3.25mm,0) {\pgfuseplotmark{*}};}}}
\newcommand{\TermTwo}{\raisebox{-1pt}{\tikz{\draw[very thick, dotted, color=ColorTwo](0,0)--(6mm,0);\node[mark size=3pt,color=ColorTwo] at (3.25mm,0) {\pgfuseplotmark{square*}};}}}
\newcommand{\TermThree}{\raisebox{-1pt}{\tikz{\draw[very thick, densely dotted, color=ColorThree](0,0)--(6mm,0);\node[mark size=3pt,color=ColorThree] at (3.25mm,0) {\pgfuseplotmark{diamond*}};}}}
\newcommand{\TermFour}{\raisebox{-1pt}{\tikz{\draw[very thick, loosely dotted, color=ColorFour](0,0)--(6mm,0);\node[mark size=3pt,color=ColorFour] at (3.25mm,0) {\pgfuseplotmark{triangle*}};}}}
\newcommand{\TermFive}{\raisebox{-1pt}{\tikz{\draw[very thick, dashed, color=ColorFive](0,0)--(6mm,0);\node[mark size=3pt,color=ColorFive] at (3.25mm,0) {\pgfuseplotmark{*}};}}}
\newcommand{\TermSix}{\raisebox{-1pt}{\tikz{\draw[very thick, densely dashdotted, color=ColorSix](0,0)--(6mm,0);\node[mark size=3pt,color=ColorSix] at (3.25mm,0) {\pgfuseplotmark{diamond*}};}}}
\newcommand{\Error}{\raisebox{-1pt}{\tikz{\draw[very thick, solid, color=red](0,0)--(6mm,0);\node[mark size=3pt,color=red] at (3.25mm,0) {\pgfuseplotmark{square*}};}}}

\shorttitle{Sparse identification of multiphase turbulence closures}
\shortauthor{S. Beetham, R. O. Fox and J. Capecelatro}

\title{Sparse identification of multiphase turbulence closures for coupled fluid--particle flows}
\author{S. Beetham\aff{1}
  \corresp{\email{snverner@umich.edu}}
 \and R. O. Fox\aff{2}
 \and  J. Capecelatro\aff{1}}

\affiliation{\aff{1}Department of Mechanical Engineering, University of Michigan,
Ann Arbor, MI 48109, USA, 
\aff{2}Department of Chemical and Biological Engineering, Iowa State University, Ames, IA, 50011, USA}

\begin{document}

\maketitle

\begin{abstract}

In this work, model closures of the multiphase Reynolds-Average Navier-–Stokes (RANS) equations are developed for homogeneous, fully-developed gas--particle flows. To date, the majority of RANS closures are based on extensions of single-phase turbulence models, which fail to capture complex two-phase flow dynamics across dilute and dense regimes, especially when two-way coupling between the phases is important. In the present study, particles settle under gravity in an unbounded viscous fluid. At sufficient mass loadings, interphase momentum exchange between the phases results in the spontaneous generation of particle clusters that sustain velocity fluctuations in the fluid. Data generated from Eulerian--Lagrangian simulations are used in a sparse regression method for model closure that ensures form invariance. Particular attention is paid to modelling the unclosed terms unique to the multiphase RANS equations (drag production, drag exchange, pressure strain and viscous dissipation). A minimal set of tensors is presented that serve as the basis for modelling. It is found that sparse regression identifies compact, algebraic models that are accurate across flow conditions and robust to sparse training data. 

\end{abstract}


\section{Introduction} 
Many natural and industrial processes involve the flow of solid particles or liquid droplets whose dynamical evolution and morphology are intimately coupled with a carrier gas. A peculiar behaviour of disperse multiphase flows is their ability to give rise to large-scale structures (hundreds to thousands of times the size of individual particles), from dense clusters to nearly particle-free voids (see figure~\ref{fig:ResultsSummary}). Such large-scale heterogeneity can effectively `demix’ the underlying flow, reducing contact between the phases resulting in enormous consequences in engineering systems \citep{shaffer2013high,miller2014carbon,Guo2019,Beetham2019}.

Seminal works by G. K. Batchelor have provided theoretical estimates describing the motion of collections of solid particles suspended in viscous flows \citep{Batchelor1972, Batchelor1982}, in addition to important insights on the instabilities present in such systems. For example, \citet{Batchelor1988} demonstrated that small rigid spheres falling under gravity will give rise to long-range hydrodynamic interactions that result in hindered settling \citep{Batchelor1972}. In more recent studies, it was demonstrated that at higher Reynolds numbers and particle concentrations, momentum exchange between the phases results in \emph{enhanced} settling when the mean mass loading, $\varphi$, defined by the ratio of the specific masses of the particle and fluid phases, is of order one or larger~\citep{capecelatro2015}. In statistically homogeneous gravity-driven gas--solid flows, the average particle settling speed, $\mathcal{V}$, can be approximated as
\begin{equation} 
\mathcal{V} = \mathcal{V}_0 + \langle u_f \rangle_p 
\end{equation}
for Stokes flow \citep{capecelatro2015}, where $\mathcal{V}_0=\tau_p g$ is the terminal Stokes settling velocity of an isolated particle with $\tau_p$ the particle response time and $g$ gravity. In this expression, the phase-averaged fluid velocity, $\langle u_f \rangle_p = \langle \alpha_p u_f \rangle / \langle \alpha_p \rangle$,  is sometimes referred to as the velocity \emph{seen} by the particles, where $u_f$ is the local fluid velocity aligned with gravity, $\alpha_p$ is the local particle volume fraction, and angled brackets denote a spatial and temporal average. At sufficient mass loading, the fluid-phase velocity and particle concentration are often highly correlated, and fluctuations in particle concentration can generate and sustain fluid-phase turbulence (as shown in figure~\ref{fig:ResultsSummary}), referred to here as cluster-induced turbulence (CIT). Because clusters entrain the carrier phase, $u_f$ and $\alpha_p$ are highly correlated, resulting in $\mathcal{V}>\mathcal{V}_0$. 

Due to the breadth of length- and time-scales present in turbulent fluid--particle mixtures, accurate modelling of industrial and environmental flows remains challenging. Thus, the Reynolds-averaged Navier--Stokes (RANS) equations are the workhorse of industry to inform engineering designs and decisions. Because of the importance of the multiphase physics present in large-scale systems, developing multiphase RANS closures that are accurate under relevant conditions is critically important.

To date, multiphase turbulence models have largely relied upon extensions to single-phase models (see e.g. \cite{sinclair1989gas}, \cite{dasgupta1994turbulent}, \cite{sundaram1994spectrum}, \cite{cao1995}, \cite{Dasgupta1998}, \cite{Cheng1999} \cite{Jiang2012} \cite{Rao2012} \cite{Zeng2006}) that were derived directly from the Navier--Stokes equations. In contrast to modelling by analogy with single-phase flow, \citet{fox2014} developed the exact Reynolds-averaged equations for collisional fluid--particle flows. In that work, it was demonstrated that directly averaging the Navier--Stokes equations fails to capture important two-phase interactions. Instead, it was demonstrated that phase averaging the mesoscale (locally averaged) equations results in a set of equations that explicitly account for two-way coupling contributions. \citet{capecelatro2015} further developed the Reynolds averaged formulation of \citet{fox2014} to include transport equations for the volume-fraction variance, drift velocity and the separate components of the Reynolds stresses of each phase and particle-phase pressure tensor. While exact, it does lead to a large number of unclosed terms that require modelling, which is the focus of the present study.

Accurate modeling of the unclosed terms that remain predictive from dilute to dense regimes remains an outstanding challenge. \citet{fox2014} proposed closures of the phase-averaged (PA) terms based largely on single-phase turbulence models without extensive validation. \cite{capecelatro2016channel2} extended these models to account for near-wall effects in particle-laden channel flows. Agreement with the turbulence statistics obtained from simulation data was found to be satisfactory at first order (e.g. PA velocities) but less so at second order (e.g. PA turbulent kinetic energy). \citet{Innocenti2019} drew upon a probability-density-function approach, along with extensions from single-phase turbulence modeling (particularly in the fluid phase), showing satisfactory agreement for statistics up to second order. However, the model was restricted to relatively dilute flows. Due to the large parameter space associated with turbulent multiphase flows, a reliable modeling approach valid across two-phase flow regimes (e.g. dilute to dense limit) remains elusive.


Broadly speaking, extracting new models and understanding of physics from data has a long history in many diverse areas of science and engineering \cite[see e.g.][]{ML_2015Jordan}. In the last decade, these data-driven techniques have been applied to turbulence modelling in several ways, including uncertainty prediction and quantification, model calibration and augmentation and the generation of entirely new models.  Several recent works have utilized machine learning (neural networks are particularly popular \cite{ML_2015Tracey, ML_2002Milano, ML_2010Lu, ML_2012Rajabi, ML_2014Duraisamy_transition, ML_2014Duraisamy_new, ML_2016Ma, Ling2016, Bode2019, Liu2019}) in order to translate large amounts of experimental or computational data into model closures. Neural networks have shown relatively exceptional performance outside the region in which they were trained. As a departure from more traditional modelling techniques, these methods are inserted modularly, as a `black box,' into an existing flow solver. Thus, while they have displayed a high level of performance on a wide range of flow conditions, the closure does not satisfy the interpretability condition necessary for making physical inferences. Further, a large number of neural network approaches attempt to augment or correct existing models. However, as discussed above, in the context of multiphase flows appropriate existing models in which to augment do not exist.  

Rather than relying on a best fit strategy, as done in neural networks, \citet{ML_2016Brunton} developed a strategy based on sparse regression that identifies the underlying functional form of the nonlinear physics by optimizing a coefficient matrix that acts upon a matrix of trial functions. While this method requires knowledge about the physics of the system under configuration (in order to make informed selections of the trial functions), it can be reasonably assumed that the modeller is not entirely naive. In fact, traditional modelling techniques have relied nearly exclusively on this notion. \citet{Beetham2020} recently extended the sparse identification framework of \citet{ML_2016Brunton} to ensure invariance of the resulting models, and demonstrated its utility in developing closed-form algebraic RANS models for a variety of single-phase flows.

In this work, the sparse identification modelling framework of \citet{Beetham2020} is employed to develop multiphase closure models for homogeneous, gravity-driven gas--solid flows. Eulerian--Lagrangian simulations are performed across a range of Archimedes numbers and volume fractions to provide training data. The terms appearing in the multiphase RANS equations recently derived in \citet{capecelatro2015} are extracted. We then build a minimally invariant basis set of tensors (i.e. a set of functional groups that serve as candidate terms in the desired model). Such basis sets are well established for single-phase turbulent flows \citep{ML_1991Speziale, Gatski1993, ML_1975Pope}, however an analogous basis has not yet been determined for multiphase flows. Using this basis and the sparse regression methodology, the compact functional form of the physics-based closures are inferred. As we consider exclusively statistically stationary and homogeneous systems, model realizability \citep{PopeText} is left for future work.
 
\section{System description} 
\subsection{Configuration under study}
In the present study, rigid spherical particles of diameter $d_p$ and density $\rho_p$ are suspended in an unbounded (triply periodic) domain containing an initially quiescent gas of density $\rho_f$ and viscosity $\nu_f$. Gravity $g$ acts in the negative $x$-direction. As particles settle, they spontaneously form clusters. Due to two-way coupling between phases, particles entrain the fluid, generating turbulence therein.  A frame of reference with the fluid phase is considered, such that the mean streamwise fluid velocity is null. Given the relative simplicity of the configuration, only a few non-dimensional groups arise. An important non-dimensional number is the Archimedes number, defined as
\begin{equation}
\Arc =  (\rho_p/\rho_f -1)d_p^3 g/\nu_f^2.
\end{equation} 
Alternatively, a Froude number can be introduced to characterize the balance between gravitational and inertial forces, defined as $\Fr= \tau_p g/ d_p$, where $\tau_p = {\rho_p d_p^2}/{(18 \rho_f \nu_f)}$ is the particle response time. The Stokes settling velocity for an isolated particle is given by $\mathcal{V}_0=\tau_p g$. From this a characteristic cluster length can also be estimated \emph{a priori} as $\mathcal{L} = \tau_p^2 g$. To ensure the hydrodynamics are independent of the domain size, the simulation configurations are equal or larger than Case 4 reported in \citet{Capecelatro2016_domain}.  

\begin{table}
\centering
\def~{\hphantom{0}}
\begin{tabular}{l @{\hskip 0.2in} l @{\hskip 0.2in} l c c c }
\multicolumn{6}{c}{\textit{Dimensional Quantities}} \\ 
$\mathcal{V}_0$ & Stokes settling velocity & [m/s] &0.02 & 0.06&0.2\\ 
$\mathcal{L}$ & Characteristic cluster length & [m] & $5.0\times10^{-4}$& $1.5\times10^{-3}$& $5.0\times10^{-3}$ \\ 
$\tau_p$ & Drag time & [s] &\multicolumn{3}{c}{0.025} \\ 
$\rho_p$ & Particle density & [kg/m$^3$] & \multicolumn{3}{c}{1000}  \\ 
$d_p$ & Particle diameter & [m] & \multicolumn{3}{c}{$90\times 10^{-6}$} \\ 
$\rho_f$ & Fluid density & [kg/m$^3$] & \multicolumn{3}{c}{1}  \\ 
$\nu_f$ & Fluid viscosity & [m$^2$/s] & \multicolumn{3}{c}{$1.8 \times 10^{-5}$}\\ 
$g$ & Gravity & [m/s$^2$] & 0.8 & 2.4 & 8.0 \\ [10pt]
\multicolumn{6}{c}{} \\
 \multicolumn{6}{c}{\textit{Non-dimensional Quantities}} \\
$N_p$ & Number of particles&  & \multicolumn{3}{c}{(610,370, 15,564,442, 30,518,514)}\\
$\langle \alpha_p \rangle$ & Mean volume fraction $\times 10^{-2}$ &  & \multicolumn{3}{c}{(0.1, 2.55, 5.0)} \\
$\varphi$ & Mean mass loading &  & \multicolumn{3}{c}{(1.0, 26.2, 52.6)}\\
\Fr & Froude number &  & 5.6 & 16.7 & 55.6 \\ 
\Arc & Archimedes number &  & 1.8 & 5.4 & 18.0\\ 
\multicolumn{6}{c}{} \\
\multicolumn{6}{c}{\textit{Computational Quantities}} \\
 \multicolumn{2}{c}{Domain size} & {[m]} & \multicolumn{3}{c}{$0.158 \times 0.038 \times 0.038$} \\ 
 \multicolumn{2}{c}{Grid size} &  & \multicolumn{3}{c}{$512 \times 128 \times 128$} \\ 
 \multicolumn{2}{c}{$L_x/\mathcal{L}$} & & \multicolumn{1}{c}{316} &  \multicolumn{1}{c}{105} & \multicolumn{1}{c}{32} \\
\end{tabular}
\caption{Summary of parameters for the configurations under consideration.} 
\label{tab:config} 
\end{table} 

To sample the parameter space typical of turbulent fluidized bed reactors~\citep{Sun2019}, the mean particle-phase volume fraction is varied from $0.001\le\langle\alpha_p\rangle\le0.05$ and the Archimedes number is varied from $1.8\le\Arc\le18.0$ by adjusting gravity. Due to the large density ratios under consideration, the mean mass loading ranges from $\mathcal{O}(10)$--$\mathcal{O}(10^2)$, and consequently two-way coupling between the phases is expected to be important. Here, angled brackets denote both a spatial and a temporal average (since the flow under consideration is triply periodic and statistically stationary in time). A list of relevant non-dimensional numbers and other important simulation parameters are summarized in table~\ref{tab:config}. 

\subsection{Volume-filtered equations}
In this section, we present the volume filtered Eulerian--Lagrangian equations used to formulate the Reynolds averaged equations in \S~\ref{sec:PA} and generate the simulation data that will be applied to the sparse regression methodology in \S~\ref{sec:closure}. The position and velocity of the $i$-th particle is calculated according to Newton's second law
\begin{equation}\label{eq:newton}
\frac{{\rm{d}} \bm{x}_p^{(i)}}{{\rm{d}}t} =\bm{v}_p^{(i)}\quad {\rm{and}}\quad
\frac{{\rm{d}} \bm{v}_p^{(i)}}{{\rm{d}}t} = \mathcal{A}^{(i)} + \bm{F}_c^{(i)} + \bm{g},
\end{equation}
where $\bm{x}_p^{(i)}$ is the centre position of particle $i$ and $\bm{v}_p^{(i)}$ is its velocity at time $t$ and $\bm{g} = (-g,\,0,\,0)^{{\rm{T}}}$ is the acceleration due to gravity. The force due to inter-particle collisions, $\bm{F}_c$, is accounted for using a soft-sphere collision model originally proposed by \citet{Cundall1979}. Particles are treated as inelastic and frictional with a coefficient of restitution of $0.85$ and coefficient of friction of $0.1$~\citep{Capecelatro2013}. Momentum exchange between the phases is given by
\begin{equation}
 \mathcal{A}^{(i)} = \frac{ {F}_d }{\tau_p}  \left(\bm{u}_f - \bm{v}_p^{(i)} \right)- \frac{1}{\rho_p} \nabla p_f + \frac{1}{\rho_p} \nabla \cdot \boldsymbol{\sigma}_f, \label{eq:Adef}
 \end{equation} 
where $\bm{u}_f$, $p_f$ and $\boldsymbol{\sigma}_f$ are the fluid-phase velocity, pressure, and viscous stress tensor evaluated at the particle location, respectively and $F_d(\alpha_f,\Reyn_p)$ is the nondimensional drag correction of \citet{Tenneti2011} that takes into account local volume fraction and Reynolds number effects
where $\alpha_f=1-\alpha_p$ is the fluid-phase volume fraction and the particle Reynolds number is defined as 
 \begin{equation}
 \Reyn_p = \frac{ \alpha_f \vert  \bm{u}_f  -  \bm{v}_p^{(i)}  \vert d_p}{\nu_f}.
 \end{equation}
$F_d(\alpha_f,\Reyn_p)$ reduces to the classical Reynolds-number drag correction of~\citep{schiller1935drag} in the limit of low volume fraction ($\alpha_f=1$) and Stokes drag when $\alpha_f=1$ and $\Reyn_p=0$.

To account for the presence of particles in the fluid phase without resolving the boundary layers around individual particles, a volume filter is applied to the incompressible Navier--Stokes equations \citep{Anderson1967}. This procedure replaces the point variables with smooth, locally filtered fields. These volume-filtered equations are given by 
\begin{equation}
\frac{\partial \alpha_f}{\partial t} + \nabla \cdot \left(\alpha_f \bm{u}_f\right) = 0 
\end{equation}
and 
\begin{equation}
\frac{\partial \alpha_f \bm{u}_f}{\partial t} + \nabla \cdot \left( \alpha_f \bm{u}_f \otimes \bm{u}_f \right) = -\frac{1}{\rho_f} \nabla p_f + \nabla \cdot \boldsymbol{\sigma}_f - \frac{\rho_p}{\rho_f} \alpha_p \mathcal{A} + \alpha_f \bm{g},
\end{equation}
where $\mathcal{A}$ is the locally averaged momentum exchange term, evaluated at each Lagrangian particle and projected to the Eulerian mesh. The fluid-phase viscous-stress tensor is defined as 
\begin{equation}
\boldsymbol{\sigma}_{f} = \nu_f \left \lbrack \nabla \bm{u}_f + \left(\nabla \bm{u}_f \right)^{{\rm{T}}} - \frac{2}{3} \nabla \cdot \bm{u}_f \bm{\mathbb{I}} \right \rbrack,
\end{equation}
where $\bm{\mathbb{I}}$ is the identity matrix.

The Eulerian--Lagrangian equations are solved using NGA~\citep{desjardins2008high}, a fully conservative, low-Mach number finite volume solver. A pressure Poisson equation is solved to enforce continuity via fast Fourier transforms in all three periodic directions. The fluid equations are solved on a staggered grid with second-order spatial accuracy and advanced in time with second-order accuracy using the semi-implicit Crank--Nicolson scheme of \citet{pierce2001progress}.  Lagrangian particles are integrated using a second-order Runge--Kutta method. Fluid quantities appearing in Eq.~\eqref{eq:newton} are evaluated at the position of each particle via trilinear interpolation. Particle data is projected to the Eulerian mesh using the two-step filtering process described in \citet{Capecelatro2013}. 

\subsection{Eulerian--Lagrangian training data \label{sec:trusted}}
\begin{figure}
\centering
 \subcaptionbox{$\Arc = 1.80$\label{Resultsa}}
     {  \begin{tabular}{c c c}
       \multicolumn{3}{c}{$\langle \alpha_p \rangle$} \\
       \hline
       $0.001$ & $0.0255$ & $0.05$ \\
       \includegraphics[height = 0.35 \textwidth ]{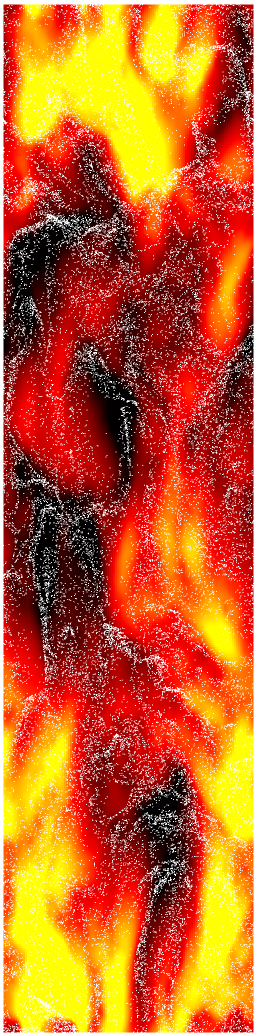} & 
       \includegraphics[height = 0.35 \textwidth ]{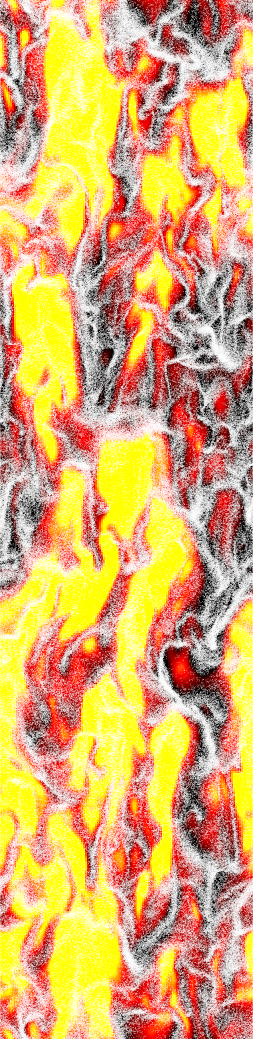} &  
       \includegraphics[height = 0.35 \textwidth]{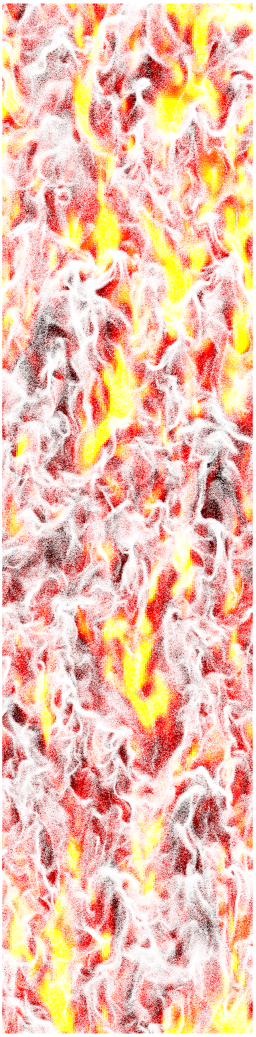} \\
       &  \includegraphics[height = 0.06 \textwidth ]{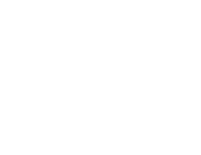} & 
 \end{tabular} }
  \subcaptionbox{$\Arc = 5.40$\label{Resultsb}}
      { \begin{tabular}{c c c}
       \multicolumn{3}{c}{$\langle \alpha_p \rangle$} \\
       \hline
       $0.001$ & $0.0225$ & $0.05$ \\
       \includegraphics[height = 0.35 \textwidth ]{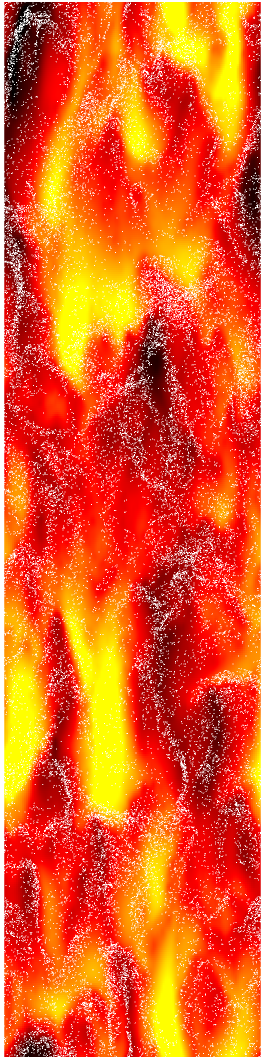} & 
       \includegraphics[height = 0.35 \textwidth ]{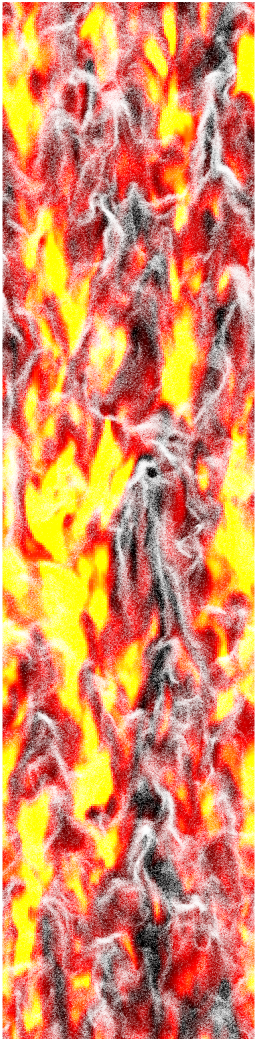} &  
       \includegraphics[height = 0.35 \textwidth]{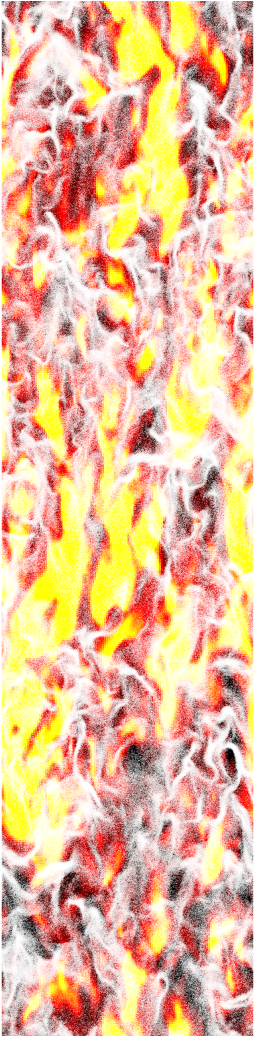} \\
        & \includegraphics[height = 0.06 \textwidth ]{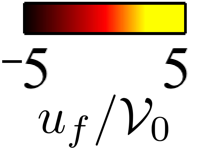} &  
 \end{tabular} }
          \subcaptionbox{$\Arc= 18.0$\label{Resultsc}}
      { \begin{tabular}{c c c}
       \multicolumn{3}{c}{$\langle \alpha_p \rangle$} \\
       \hline
       $0.001$ & $0.0255$ & $0.05$ \\
       \includegraphics[height = 0.35 \textwidth ]{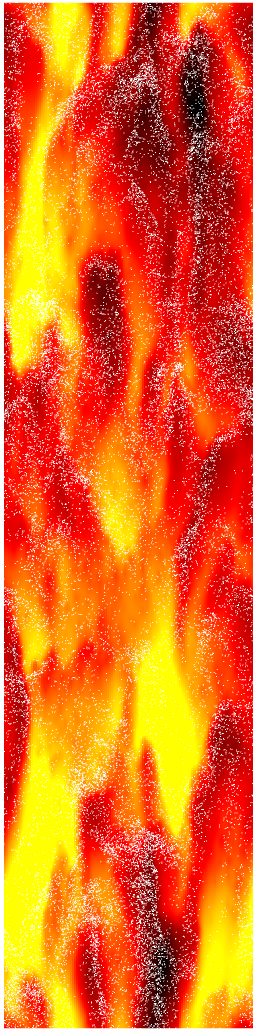} & 
       \includegraphics[height = 0.35 \textwidth ]{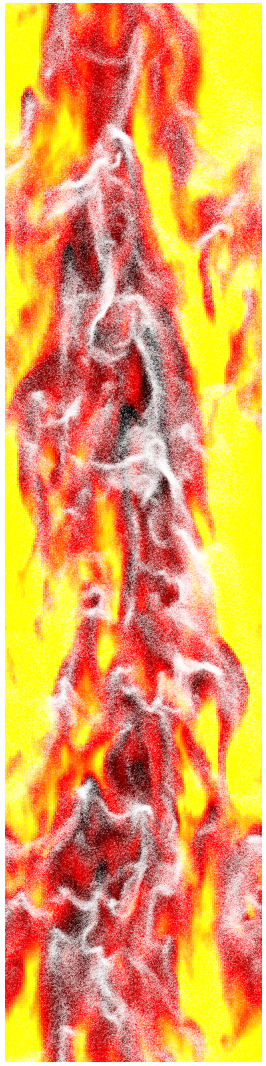} &  
       \includegraphics[height = 0.35 \textwidth]{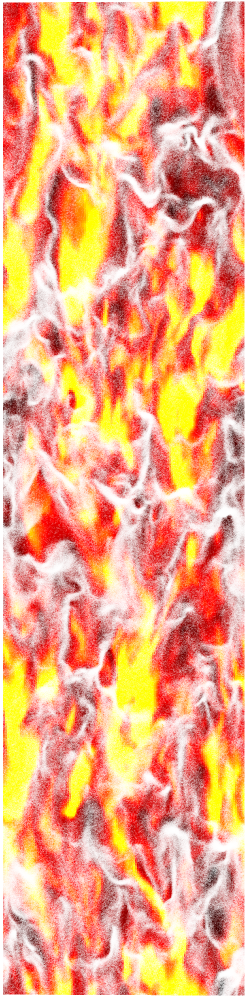} \\
       &  \includegraphics[height = 0.06 \textwidth ]{spacer} & 
 \end{tabular} }\\
\caption{Instantaneous snapshots of fully-developed CIT at statistical steady state. A slice at the centreline in the $x$--$y$ plane is shown, with particle position (white) and normalized vertical fluid velocity $u_f/\mathcal{V}_0$ (colour).}
\label{fig:ResultsSummary}  
\end{figure}

The Eulerian--Lagrangian simulations were initialized with a random distribution of particles and run for approximately $100 \tau_p$ until the flow reached a statistically stationary state. At this point statistics are accumulated over $50 \tau_p$. Instantaneous snapshots of the streamwise fluid velocity and particle position of each case at steady state are shown in figure~\ref{fig:ResultsSummary}. It can immediately be seen that clusters of particles are generated and entrain the fluid downward. As a consequence of the frame of reference under consideration, the fluid flows upward in regions void of particles. Clusters are seen to become more distinct with increasing $\langle\alpha_p\rangle$. The effect of $\Arc$ on the flow field is less noticeable. As shown in table~\ref{tab:tstar}, the standard deviation in volume fraction fluctuations $\langle {\alpha_p^\prime}^2\rangle^{1/2}$ increases with increasing $\Arc$, with $\alpha_p^\prime=\alpha_p-\langle\alpha_p\rangle$, indicating enhanced clustering. Perhaps less obvious, the volume fraction fluctuations normalized by $\langle\alpha_p\rangle$ are maximum for the intermediate volume fraction case ($\langle\alpha_p\rangle=0.025$).

\begin{table}
\centering
\begin{tabular}{l @{\hskip 0.1in} l @{\hskip 0.2in} c @{\hskip 0.2in} c  c c c @{\hskip 0.2in} c  c} 
&$\langle \alpha_p \rangle$ & $\frac{\sqrt{\langle \alpha_p^{\prime 2} \rangle}}{\langle \alpha_p \rangle}$ & $\frac{\langle u_f^{\prime \prime \prime 2} \rangle_f }{k_f}$ & $\frac{\langle v_f^{\prime \prime \prime 2} \rangle_f}{k_f}$ & $\frac{\langle u_p^{\prime \prime 2} \rangle_p}{k_p}$ & $\frac{\langle v_p^{\prime \prime 2} \rangle_p}{k_p}$ & $\frac{\langle u_p \rangle_p}{\mathcal{V}_0}$ & $ \frac{\left(\langle u_f \rangle_p + \tau_p^{\star} g\right)}{\mathcal{V}_0}$ \\
& & & & & & & & \\ 
& 0.001 & 0.63 & 1.49 & 0.25 & 1.48 & 0.26 & 1.87 & 1.53 \\
\Arc = 1.8& 0.0255&0.76 & 1.58 & 0.21 & 1.61 & 0.20 & 2.56 & 2.39  \\
 & 0.05 & 0.74&  1.51 & 0.24 & 1.52 & 0.24 & 2.46 & 2.34  \\
 & & & & & & & & \\ 
 & 0.001 & 0.71& 1.68 & 0.16 & 1.67 & 0.16 & 1.63 & 1.36 \\
\Arc = 5.4& 0.0225 & 0.87& 1.61 & 0.19 & 1.63 & 0.19 & 2.28 & 2.21 \\
 & 0.05 & 0.84& 1.56 & 0.22 & 1.57 & 0.21 & 2.12 & 2.07  \\
& & & & & & & & \\ 
   & 0.001 & 0.72&  1.76 & 0.12 & 1.76 & 0.12 & 1.26 & 1.15 \\
\Arc = 18.0& 0.0255& 1.00& 1.70 & 0.15 & 1.73 & 0.14 & 1.86 & 1.91 \\
 & 0.05 & 0.98& 1.64 & 0.18 & 1.68 & 0.16& 1.81 & 1.83 \\
\end{tabular}
\caption{Statistically stationary EL quantities for all nine training cases.}
\label{tab:tstar}
\end{table}

\section{Phase-averaged equations} \label{sec:PA}
In this section, we present the phase-averaged flow equations in which we seek to model the unclosed terms that arise. This system of equations have been previously derived \citep{capecelatro2015} and is extended here to take into account nonlinear drag effects due to $F_d$ in Eq.~\eqref{eq:Adef}. Phase averaging (PA) is analogous to Favre averaging of variable-density flows and is denoted by $\langle (\cdot) \rangle_p = \langle \alpha_p (\cdot) \rangle/\langle \alpha_p \rangle$. Fluctuations about the PA particle velocity are expressed as $\bm{u}_p^{\prime \prime} = \bm{u}_p(\bm{x},t) - \langle \bm{u}_p \rangle_p$, with $\langle \bm{u}_p^{  \prime \prime} \rangle_p = 0$. This gives rise to the PA particle-phase turbulent kinetic energy (TKE), $k_p =  \langle \bm{u}_p^{\prime \prime} \cdot  \bm{u}_p^{\prime \prime} \rangle_p/2$. Here, $\bm{u}_p$ is the Eulerian particle-phase velocity. It should be noted that $\langle\bm{u}_p\rangle_p$ is equivalent to the average particle velocity $\langle\bm{v}_p\rangle$ (with angled brackets here used to represent a particle average). Thus, $\langle u_p\rangle_p$ will be used throughout to characterize the mean settling velocity of the particle phase.

In a similar fashion, the PA operator in the fluid phase is defined as $\langle (\cdot) \rangle_f = \langle \alpha_f (\cdot) \rangle/\langle \alpha_f \rangle$. Fluctuations about the PA fluid velocity are given by $\bm{u}_f^{\prime \prime \prime} = \bm{u}_f(\bm{x},t) - \langle \bm{u}_f \rangle_f$. With this, the fluid-phase TKE is given by $k_f = \langle \bm{u}_f^{\prime \prime \prime} \cdot  \bm{u}_f^{\prime \prime \prime} \rangle_f/2$. 

The Reynolds-averaged fluid-phase equations are given by the continuity equation,
\begin{equation}
\frac{\partial \langle \alpha_f \rangle}{\partial t} + \nabla \cdot \langle \alpha_f \rangle \langle \bm{u}_f \rangle_f = 0 
\end{equation}
which is closed, and the momentum equation
\begin{align}
\frac{\partial \left (\langle \alpha_f \rangle \langle \bm{u}_f \rangle_f \right)}{\partial t} + &\nabla \cdot \langle \alpha_f \rangle \left( \langle \bm{u}_f \rangle_f \otimes \langle \bm{u}_f \rangle_f + \langle \bm{u}_f^{\prime \prime \prime} \otimes \bm{u}_f^{\prime \prime \prime} \rangle_f \right) = \nonumber \\
& \frac{1}{\rho_f} \left( \nabla \cdot \langle \boldsymbol{\sigma_f} \rangle - \nabla \langle p_f \rangle \right) - \langle \alpha_f \rangle \varphi \langle \mathcal{A} \rangle_p + \langle \alpha_f \rangle \bm{g},
\end{align}
where $\varphi=\rho_p \langle \alpha_p \rangle/ (\rho_f \langle \alpha_f \rangle)$ is the mean mass loading. The volume-filtered momentum equation results in three unclosed terms: the fluid-phase Reynolds stress tensor, $\langle \bm{u}_f^{\prime \prime \prime} \otimes \bm{u}_f^{\prime \prime \prime} \rangle_f $; the averaged fluid-phase viscous stress tensor, $\langle \boldsymbol{\sigma}_f \rangle$; and the PA interphase exchange term $\langle\mathcal{A}\rangle_p$.  The nonlinear drag correction is decomposed into $F_d = \langle {F}_d \rangle_p + F_d^{\prime \prime}$, which yields 
\begin{equation}
\langle \mathcal{A} \rangle_p = \frac{1}{\tau_p^{\star}} \left( \langle \bm{u}_f \rangle_p - \langle \bm{u}_p\rangle_p\right) - \frac{1}{\rho_p} \nabla \langle p_f \rangle_p + \frac{1}{\rho_p} \nabla \cdot \langle \boldsymbol{\sigma}_f \rangle_p.
\end{equation}
Here, we incorporate the nonlinearities associated with drag in $\tau_p^{\star} = \tau_p/\langle F_d \rangle_p$, where $\langle F_d \rangle_p(\langle\alpha_f\rangle,\langle\Reyn_p\rangle)$ is the nonlinear drag correction of \citet{Tenneti2011} with averaged flow arguments. This definition does not include the dependencies on drag covariance terms (i.e. $\langle u_f^{\prime \prime \prime} F_d^{\prime \prime}\rangle_p$ and $\langle u_p^{\prime \prime} F_d^{\prime \prime}\rangle_p$), however, as shown in table~\ref{tab:tstar}, the contributions from the drag covariance terms are negligible for describing particle settling, $\langle u_p \rangle_p$, and are thus neglected.  

For the statistically stationary and homogeneous flows considered herein, continuity implies $\langle\alpha_f\rangle$ is constant and the fluid-phase momentum yields $\langle\bm{u}_f\rangle_f=0$. In the particle phase, the only non-zero component of the averaged momentum equation is in the gravity-aligned direction (the $x$-direction in this case): 
\begin{align}
\frac{\partial \langle u_p \rangle_p}{\partial t} = \frac{1}{\tau_p^{\star}}\left(  \langle u_f \rangle_p - \langle u_p\rangle_p \right) + \frac{1}{\rho_p} \left( \left \langle \frac{\partial \sigma_{f,xi}}{\partial x_i} \right \rangle_p - \left \langle \frac{\partial p_f}{\partial x} \right \rangle_p \right) + g,
\end{align}
noting that for gas--solid flows, the terms involving $\sigma_{f,xi}$ and $p_f$ are small enough to be neglected \citep{capecelatro2015}. This implies that at steady state, $\langle u_p\rangle_p \approx \langle u_f \rangle_p + \tau_p^{\star} g$.  

The transport equations for the fluid-phase Reynolds stresses can be reduced to two unique, non-zero components. In the streamwise direction this equation is given as
\begin{align}
\label{eq:PA_ufuf}
\frac{1}{2} \frac{\partial \langle u_f^{\prime \prime \prime 2}\rangle_f}{\partial t } = &\underbrace{\frac{1}{\rho_f} \left \langle p_f \frac{\partial \langle u_f^{\prime \prime \prime} \rangle}{\partial x}\right \rangle }_{\text{\scriptsize{pressure strain (PS)}}} -  \underbrace{\frac{1}{\rho_f} \left \langle \sigma_{f,1i} \frac{\partial \langle u_f^{\prime \prime \prime} \rangle }{\partial x}\right \rangle }_{\text{\scriptsize{viscous dissipation (VD)}}} + \underbrace{ \frac{\varphi}{\tau_p^{\star}}\left( \langle u_f^{\prime \prime \prime} \rangle \langle u_p^{\prime\prime} \rangle_p - \langle u_f^{\prime \prime \prime 2}\rangle_p \right)}_{\text{\scriptsize{drag exchange (DE)}}} + \nonumber \\
&\underbrace{\frac{\varphi}{\tau_p^{\star}} \langle u_f^{\prime \prime \prime}\rangle \langle u_p\rangle_p }_{\text{\scriptsize{drag production (DP)}}} + \underbrace{\frac{\varphi}{\rho_p} \left \langle u_f^{\prime\prime\prime}\frac{\partial p^{\prime}_f}{\partial x} \right \rangle_p }_{\text{\scriptsize{pressure exchange (PE)}}} - \underbrace{\frac{\varphi}{\rho_p} \left \langle  u_f^{\prime \prime \prime} \frac{\partial \sigma^{\prime}_{f,1i}}{\partial x_i} \right \rangle_p }_{\text{\scriptsize{viscous exchange (VE)}}}.
\end{align}
Similarly, both cross-stream equations are given as
\begin{align}
\label{eq:PA_vfvf}
\frac{1}{2} \frac{\partial \langle v_f^{\prime \prime \prime 2}\rangle_f}{\partial t } = &\underbrace{\frac{1}{\rho_f} \left \langle p_f \frac{\partial \langle v_f^{\prime \prime \prime}\rangle}{\partial y}\right \rangle }_{\text{\scriptsize{pressure strain (PS)}}} -  \underbrace{\frac{1}{\rho_f} \left \langle \sigma_{f,2i} \frac{\partial \langle v_f^{\prime \prime \prime}\rangle}{\partial x}\right \rangle }_{\text{\scriptsize{viscous dissipation (VD)}}} + \underbrace{ \frac{\varphi}{\tau_p^{\star}}\left( \langle v_f^{\prime \prime \prime} v_p^{\prime\prime} \rangle_p - \langle v_f^{\prime \prime \prime 2}\rangle_p \right)}_{\text{\scriptsize{drag exchange (DE)}}} + \nonumber \\
& \underbrace{\frac{\varphi}{\rho_p} \left \langle v_f^{\prime\prime\prime}\frac{\partial p^{\prime}_f}{\partial y} \right \rangle_p }_{\text{\scriptsize{pressure exchange (PE)}}} - \underbrace{\frac{\varphi}{\rho_p} \left \langle  v_f^{\prime \prime \prime} \frac{\partial \sigma^{\prime}_{f,2i}}{\partial x_i} \right \rangle_p }_{\text{\scriptsize{viscous exchange (VE)}}},
\end{align}
where the drag production term no longer appears, since it is a gravity-driven phenomenon.

Due to the homogeneity of the flow and symmetry in the directions perpendicular to gravity ($y$ and $z$ directions in this configuration), the unique, non-zero PA Reynolds-stress transport equations in the particle phase are given as
\begin{align}
\frac{1}{2} \frac{\partial \langle u_p^{\prime \prime 2} \rangle_p}{\partial t} = & \underbrace{\left \langle \Theta \frac{\partial u_p^{\prime \prime}}{\partial x} \right \rangle_p}_{\text{\scriptsize{pressure strain}}} - \underbrace{\left \langle \sigma_{p,1i} \frac{\partial u_p^{\prime \prime}}{\partial x_i} \right \rangle_p }_{\text{\scriptsize{viscous dissipation}}}+ \underbrace{\frac{1}{\tau_p^{\star}} \left( \langle u_f^{\prime \prime \prime} u_p^{\prime \prime} \rangle_p - \langle u_p^{\prime \prime 2} \rangle_p\right)}_{\text{\scriptsize{drag exchange}}} \nonumber \\
& \underbrace{\frac{1}{\rho_p}\left \langle u_p^{\prime \prime} \frac{\partial \sigma_{f,1i}^{\prime}}{\partial x_i} \right \rangle_p}_{\text{\scriptsize{viscous exchange}}} - \underbrace{\frac{1}{\rho_p} \left \langle u_p^{\prime \prime} \frac{\partial p'_f}{\partial x}\right \rangle_p }_{\text{\scriptsize{pressure exchange}}}.
\end{align}
Similarly, the cross-gravity equations are both (due to symmetry and homogeneity) given as
\begin{align}
\frac{1}{2} \frac{\partial \langle v_p^{\prime \prime 2} \rangle_p}{\partial t} = & \underbrace{\left \langle \Theta \frac{\partial v_p^{\prime \prime}}{\partial y} \right \rangle_p}_{\text{\scriptsize{pressure strain}}} - \underbrace{\left \langle \sigma_{p,2i} \frac{\partial v_p^{\prime \prime}}{\partial x_i} \right \rangle_p }_{\text{\scriptsize{viscous dissipation}}}+ \underbrace{\frac{1}{\tau_p^{\star}} \left( \langle v_f^{\prime \prime \prime} v_p^{\prime \prime} \rangle_p - \langle v_p^{\prime \prime 2} \rangle_p\right)}_{\text{\scriptsize{drag exchange}}} \nonumber \\
& \underbrace{\frac{1}{\rho_p}\left \langle v_p^{\prime \prime} \frac{\partial \sigma_{f,2i}^{\prime}}{\partial y_i} \right \rangle_p}_{\text{\scriptsize{viscous exchange}}} - \underbrace{\frac{1}{\rho_p} \left \langle v_p^{\prime \prime} \frac{\partial p'_f}{\partial y}\right \rangle_p }_{\text{\scriptsize{pressure exchange}}},
\end{align}
where $\Theta$ and $\sigma_{p}$ are the granular temperature and the particle-phase viscous stress tensor, respectively \citep{capecelatro2015}. 

\section{Closure modelling}\label{sec:closure}
\subsection{Sparse regression with embedded invariance}\label{sec:closure1}
The focus of this section is modelling the unclosed terms that appear in the fluid-phase Reynolds-stress equations \eqref{eq:PA_ufuf} and \eqref{eq:PA_vfvf}. The data used to inform these closures, as discussed in \S~\ref{sec:trusted}, is averaged after the flow has become statistically stationary in time. These values are summarized in the table~\ref{tab:RSfluid}. In the streamwise direction, drag production (DP) is mostly balanced by drag exchange (DE). Pressure strain (PS) and viscous dissipation (VD) contain fluid-phase residual contributions, while pressure exchange (PE) and viscous exchange (VE) contain contributions from both phases. These terms are small compared to DP and DE, but are not negligible in general. In the cross-stream direction, DE is mostly balanced by PS.
\begin{table}
\centering
\begin{tabular}{c c  @{\hskip 0.2in} c c c c c c}
\multicolumn{8}{c}{\textit{streamwise direction}} \\
&$\langle \alpha_p \rangle $ &PS & VD & DP & DE &VE & PE  \\
\multirow{3}{*}{$\Arc = 1.8$} & {0.0100} &-0.004& -0.007& 0.03& -0.02 &0  &0\\ 
                                            & {0.0255} &-0.189 & -0.122& 2.49& -2.75 & -0.01 & 0 \\ 
                                            & {0.5000} & -0.406 & -0.162 & 6.04& -6.87 & -0.04 & 0\\ 
                                            \vspace{0.25em}\\
\multirow{3}{*}{$\Arc = 5.4$} & {0.0100} &-0.021 & -0.049& 0.15 & -0.138 & 0 & 0 \\ 
                                            & {0.0255} &-1.201 & -0.482 & 15.18 & -15.71 & -0.06 & 0\\ 
                                            & {0.5000} & -2.680 & -0.709 & 39.67 & -43.00 & -0.29 & 0\\ 
                                             \vspace{0.25em}\\
\multirow{3}{*}{$\Arc = 18.0$} & {0.0100} & -0.106 & -0.264 & 0.81 & -0.732  & 0 & 0 \\ 
                                            & {0.0255} & -9.988 & -2.097& 129.99 & -131.1 & -0.83 & 0 \\ 
                                            & {0.5000} & -22.056 & -3.455& 317.84 & -329.3 &-3.32 & 0\\ 
\multicolumn{8}{c}{} \\
\multicolumn{8}{c}{\textit{cross-stream directions}} \\
 &$\langle \alpha_p \rangle $ &PS & VD & DE  &VE & PE  \\
\multirow{3}{*}{$\Arc = 1.8$} & {0.0100} & 0.002& -0.0004& -0.002&0 &0 \\ 
                                            & {0.0255} &0.096 & -0.0079& -0.171&0.002&0\\ 
                                            & {0.5000} &0.219 & -0.0140& -0.455 & 0.012 &0.002  \\ 
                                             \vspace{0.25em}\\
\multirow{3}{*}{$\Arc = 5.4$} & {0.0100} &0.011 & -0.0016 & -0.01 & 0&0 \\ 
                                            & {0.0225} &0.616  & -0.0293& -0.83 & 0.017& 0.002 \\ 
                                            & {0.5000} & 1.425 & -0.0537 & -2.323 & 0.083 & 0.007 \\ 
                                             \vspace{0.25em}\\
\multirow{3}{*}{$\Arc = 18.0$} & {0.0100}& 0.053 & -0.0052 & -0.047 & 0 & 0 \\  
                                            & {0.0255} & 5.144 & -0.1102& -5.686&  0.183 & 0.009 \\ 
                                            & {0.5000} & 12.020 & -0.2305 & -17.637 & 0.672&0.051 \\ 
\end{tabular}
\caption{Averaged terms for each contribution in the fluid-phase Reynolds-stress transport equations \eqref{eq:PA_ufuf} and \eqref{eq:PA_vfvf}.}
\label{tab:RSfluid}
\end{table}

Each unclosed term is considered individually and models are learned using the sparse regression methodology described in \citet{Beetham2020} and summarized here. In this method, it is postulated that any tensor quantity $\mathbb{D}$ can be modelled using an invariant tensor basis, $\mathbb{T},$ and a set of ideal, sparse coefficients, $\hat{\beta}$, 
\begin{equation}
\mathbb{D} = \mathbb{T} \hat{\beta}.
\end{equation}
 The ideal coefficients are determined by solving the optimization problem
 \begin{equation}
 \hat{\beta} = \min_{\beta} \vert \vert \mathbb{D} - \mathbb{T}\mathbb{B} \vert \vert^2_2 + \lambda \vert \vert \beta \vert \vert_1,
 \end{equation}
where $\beta$ is a vector of coefficients that varies depending upon the choice of a user-specified sparsity parameter, $\lambda$ and $\vert \vert \cdot \vert \vert_2^2$ and $\vert \vert \cdot \vert \vert_1$  represent the L-2 and L-1 norms, respectively. 
In the case of single-phase turbulence, this methodology can be used readily with previously derived minimally invariant basis sets \citep{Beetham2020}. However, to date an analogous basis has not yet been identified for multiphase flows. Due to the relative simplicity of the system under study (i.e. symmetry, homogeneity and stationarity), the parameters that may contribute to such a basis are limited to three tensors: the fluid-phase Reynolds stress anisotropy tensor, $\b$, the particle-phase Reynolds stress anisotropy tensor $\c$, and a the mean slip tensor, $\a$ (see table~\ref{tab:RANStensors}). The mean slip tensor is defined as $\bm{U}_r = \bm{u}_r \otimes \bm{u}_r$, where $\bm{u}_r = \langle \bm{u}_p \rangle_p  - \langle \bm{u}_f \rangle_f$ is the slip velocity vector. An important property of this vector is that in fully developed CIT it is always aligned with the direction of the body forcing (in this case gravity).    

\begin{table}
\centering
\begin{tabular}{l l}
(1) Particle-phase anisotropic stress tensor & $\hat{\mathbb{R}}_p =  \frac{\langle \bm{u}_p^{\prime \prime \prime} \bm{u}_p^{\prime \prime \prime} \rangle}{2 k_p} - \frac{1}{3}\bm{\mathbb{I}}$ \\
(2) Fluid-phase anisotropic stress tensor & $\hat{\mathbb{R}}_f =  \frac{\langle \bm{u}_f^{\prime \prime \prime} \bm{u}_f^{\prime \prime \prime} \rangle_f}{2 k_f} - \frac{1}{3}\bm{\mathbb{I}}$ \\
(3) Slip velocity tensor & $\a = \frac{\bm{U}_r}{{\rm{tr}}\left(\bm{U}_r\right)} - \frac{1}{3} \bm{\mathbb{I}}$, \\
\end{tabular} 
\caption{Second-order, symmetric, deviatoric tensors available to the multiphase RANS equations for modelling.}
\label{tab:RANStensors}
\end{table}

Because the sparse regression methodology postulates the model to be a linear combination of the basis tensors, this implies that the basis tensors must take on the same properties as the quantity to be modelled. The four terms under consideration here are all symmetric and thus the basis tensors must also be symmetric. The three tensor quantities shown in table~\ref{tab:RANStensors} are used in order to formulate a minimally invariant basis by following the procedure described in \citet{Spencer1958}. This set of tensors, along with six scalar invariants, denoted $\mathcal{S}^{(i)}$, by definition can exactly describe the Eulerian--Lagrangian data. In the context of the sparse regression methodology, the ideal coefficients $\hat{\beta}$ may be constants or nonlinear functions of the scalar invariants, $\mathcal{S}^{(i)}$. 

\begin{table} 
\centering 
\begin{tabular}{l @{\hskip -0.001in} l @{\hskip 0.5in} l @{\hskip -0.001in} l}
$\mathcal{T}^{(1)}$ & $= \bm{\mathbb{I}}$ & $\mathcal{T}^{(2)}$ & $= \a$ \\
$\mathcal{T}^{(3)}$ & $= \a^2$ & $\mathcal{T}^{(4)}$ & $= \left(\a\b\right)^{\dagger}$ \\
$\mathcal{T}^{(5)}$ & $= \left(\a^2\b\right)^{\dagger} $ & $\mathcal{T}^{(6)}$ & $= \left(\a^2\b^2\right)^{\dagger}$ \\
$\mathcal{T}^{(7)}$ & $= \left(\a\b\c\right)^{\dagger} $ & $\mathcal{T}^{(8)}$ & $=\left( \a^2\b\c\right)^{\dagger} $ \\
$\mathcal{T}^{(9)}$ & $= \left(\b\a^2\c\right)^{\dagger}$ & $\mathcal{T}^{(10)}$ & $= \left(\a^2\b^2\c\right)^{\dagger}$ \\
$\mathcal{T}^{(11)}$ & $= \left(\a\b\a^2\c \right)^{\dagger}$ & $\mathcal{T}^{(12)}$ & $= \left(\a\b\c\a^2 \right)^{\dagger}$ \\
$\mathcal{T}^{(13)}$ & $= \b$ & $\mathcal{T}^{(14)}$ & $= \b^2$ \\
$\mathcal{T}^{(15)}$ & $= \c$ & $\mathcal{T}^{(16)}$ & $= \c^2$ \\
$\mathcal{T}^{(17)}$ & $=  \left(\a\c\right)^{\dagger}$ & $\mathcal{T}^{(18)}$ & $= \left(\a^2\c \right)^{\dagger}$ \\
$\mathcal{T}^{(19)}$ & $= \left(\a\c^2\right)^{\dagger}$ & $\mathcal{T}^{(20)}$ & $= \left(\a^2\c^2\right)^{\dagger}$ \\
$\mathcal{T}^{(21)}$ & $= \left(\b\c \right)^{\dagger}$ & $\mathcal{T}^{(22)}$ & $= \left(\b^2\c \right)^{\dagger}$ \\
$\mathcal{T}^{(23)}$ & $= \left(\b\c^2 \right)^{\dagger}$ & $\mathcal{T}^{(24)}$ & $=  \left(\b^2\c^2 \right)^{\dagger}$ \\
\end{tabular} 
\begin{tabular}{l @{\hskip -0.001in} l @{\hskip 0.5in} l @{\hskip -0.001in} l @{\hskip 0.5in} l @{\hskip -0.001in} l}
$\mathcal{S}^{(1)}$ & $= {\rm{tr}}\left(\a \b^2 \c^2 \right)$ & 
$\mathcal{S}^{(2)}$ & $= {\rm{tr}}\left( \a \b \c^2 \right)$ &
$\mathcal{S}^{(3)}$ & $= {\rm{tr}}\left( \a \b \c \right)$  \\
$\mathcal{S}^{(4)}$ & $=\Arc$ & $\mathcal{S}^{(5)}$ & $=\varphi$ & $\mathcal{S}^{(6)}$ & $=\langle \alpha_p \rangle$
\end{tabular}
\caption{Minimally invariant set of basis tensors and associated scalar invariants. Here, $\left( \cdot \right)^{\dagger}=\left( \cdot \right)+\left( \cdot \right)^T$ denotes the tensor quantity added with its transpose.}
\label{tab:basis1} 
\end{table}

\subsection{Results and discussion \label{sec:results}}
Using the set of basis tensors defined in \S~\ref{sec:closure1}, the sparse regression methodology is employed to identify closures for the terms appearing in the fluid-phase Reynolds-stress equations \eqref{eq:PA_ufuf} and \eqref{eq:PA_vfvf}, based upon the Eulerian--Lagrangian data described in \S~\ref{sec:trusted}. Since flow data is homogeneous in all three spatial directions and we consider time-averaged data, each case is zero-dimensional (i.e. a single value). 

As seen in table~\ref{tab:RSfluid}, the contributions from viscous and pressure exchange are either null, or relatively small even as mass loading is increased. For this reason, modelling efforts are directed toward the four remaining terms: drag production, pressure strain, viscous dissipation and drag exchange. Each is modelled separately, beginning with drag production as it is the sole source of fluid-phase turbulent kinetic energy in the absence of mean shear. As seen in Eq.~\eqref{eq:PA_ufuf}, it is proportional to $\langle u_f^{\prime \prime \prime} \rangle_p$, which is zero in the absence of particles.  

\begin{figure}
\centering
\includegraphics[width = 0.45\textwidth]{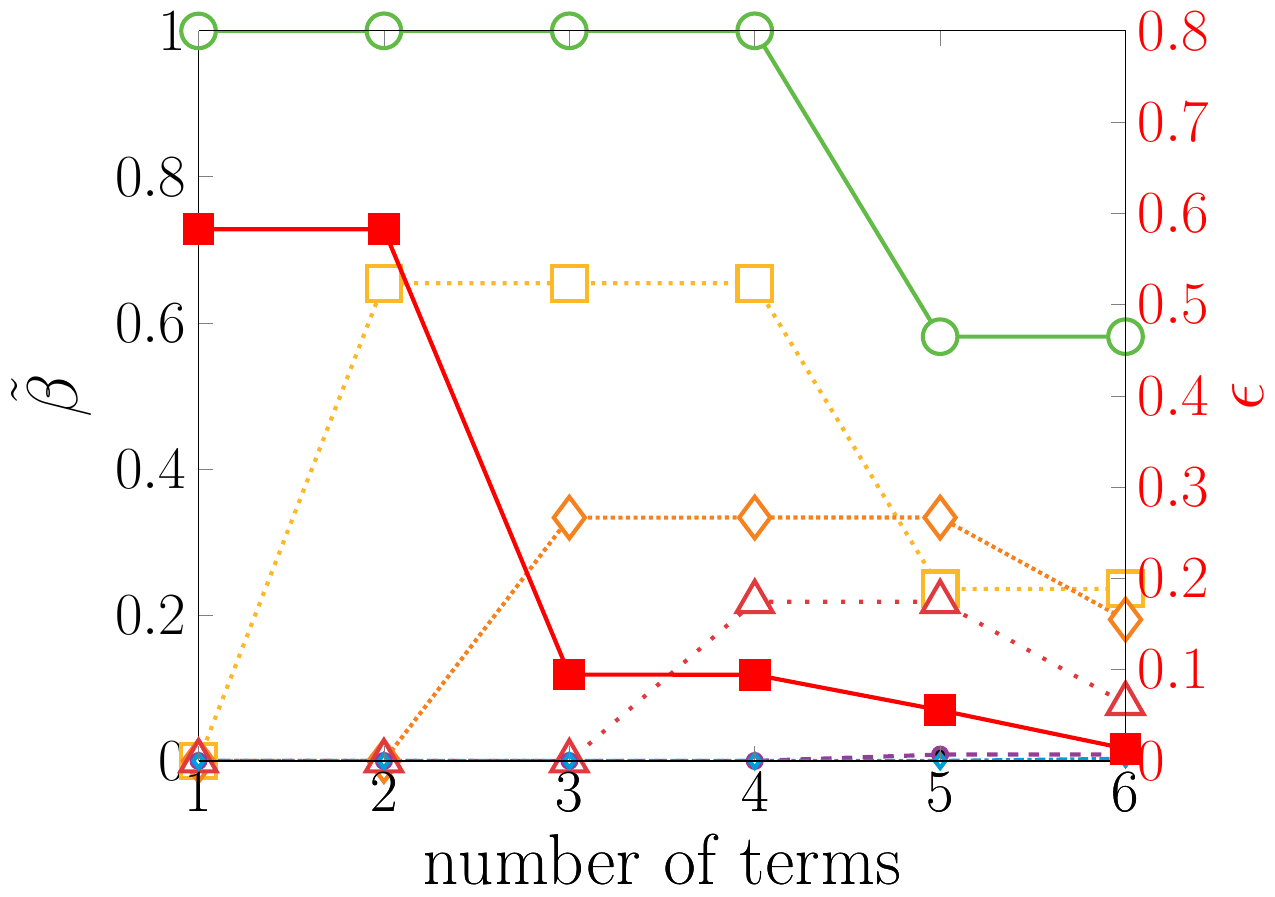}
\caption{Normalized coefficients, $\tilde{\beta}$ (left axis) and associated model error, $\epsilon$, (\protect \Error, right axis) for drag production. The two-term and five-term models are described in equations \ref{eq:DPmodelSimpler} and \ref{eq:DPmodel}, respectively. Terms 1--6 are represented as \protect \TermOne, \protect \TermTwo, \protect \TermThree, \protect \TermFour, \protect \TermFive, \protect \TermSix, respectively. These colors also correspond with figure~\ref{fig:DPcostwalk}.}
\label{DPError}
\end{figure}

As input to the sparse regression algorithm, drag production is non-dimensionalized using the square of the PA particle velocity, $\langle u_p \rangle_p^2$ and the drag time, $\tau_p$. Because drag production is symmetric and also contains zero off-diagonal components, the basis set was restricted to only include terms that are functions of $\a$ and $\mathbb{I}$, which also exhibit this property. While the Reynolds stresses have null off-diagonal components for this particular configuration, this does not hold in a general sense. 

During optimization, as $\lambda$ is decreased, additional terms are added to the learned model and model error decreases (see figure~\ref{DPError}), where model error is defined as
\begin{equation}
\epsilon = \frac{\vert \vert \mathbb{D} - \mathbb{T}\hat{\beta} \vert \vert^2_2}{\vert \vert \mathbb{D}\vert \vert^2_2}.
\end{equation}
In examining the relationship between model error and model complexity, we observe that a significant reduction in error is achieved with three model terms and error is drastically reduced when considering a six-term model. It is also notable that the most important terms to overall model performance appear in the models with lesser complexity and remain dominant as subsequent terms are added. This is indicated by the behaviour of the normalized coefficients $\tilde{\beta}$, given as $\hat{\beta}^{(p)}/\max \hat{\beta}^{(1)}$, where $p$ denotes the number of terms in the model. 

\begin{figure}
\centering
\vspace{2em}
Three-term model \\
\vspace{1em}
 \subcaptionbox{$\Arc = 1.80$\label{DP1Simpler}}
     { \includegraphics[height = 0.24\textwidth]{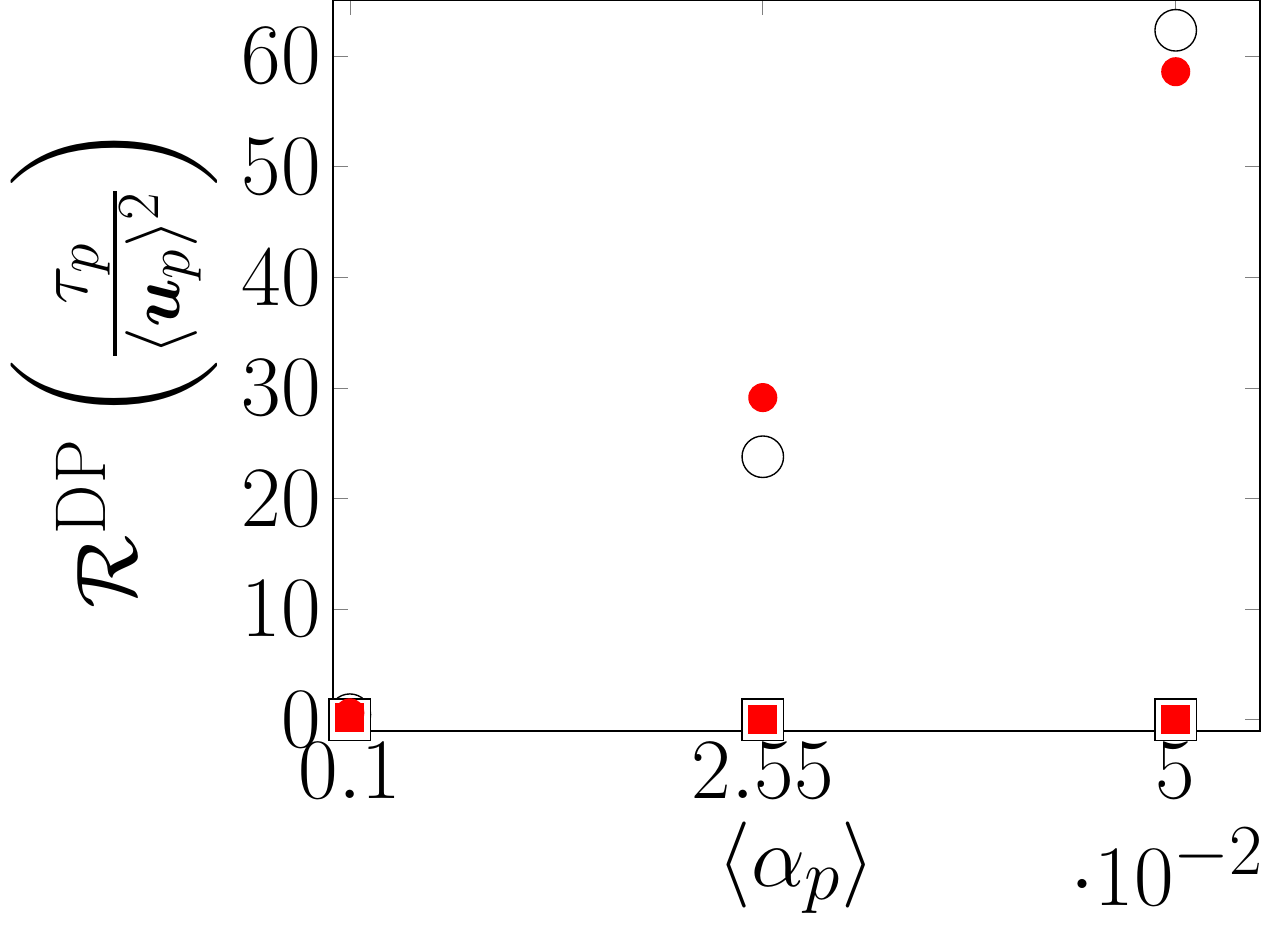}
      }
 \subcaptionbox{$\Arc = 5.40$\label{DP2Simpler}}
     { \includegraphics[height = 0.24\textwidth]{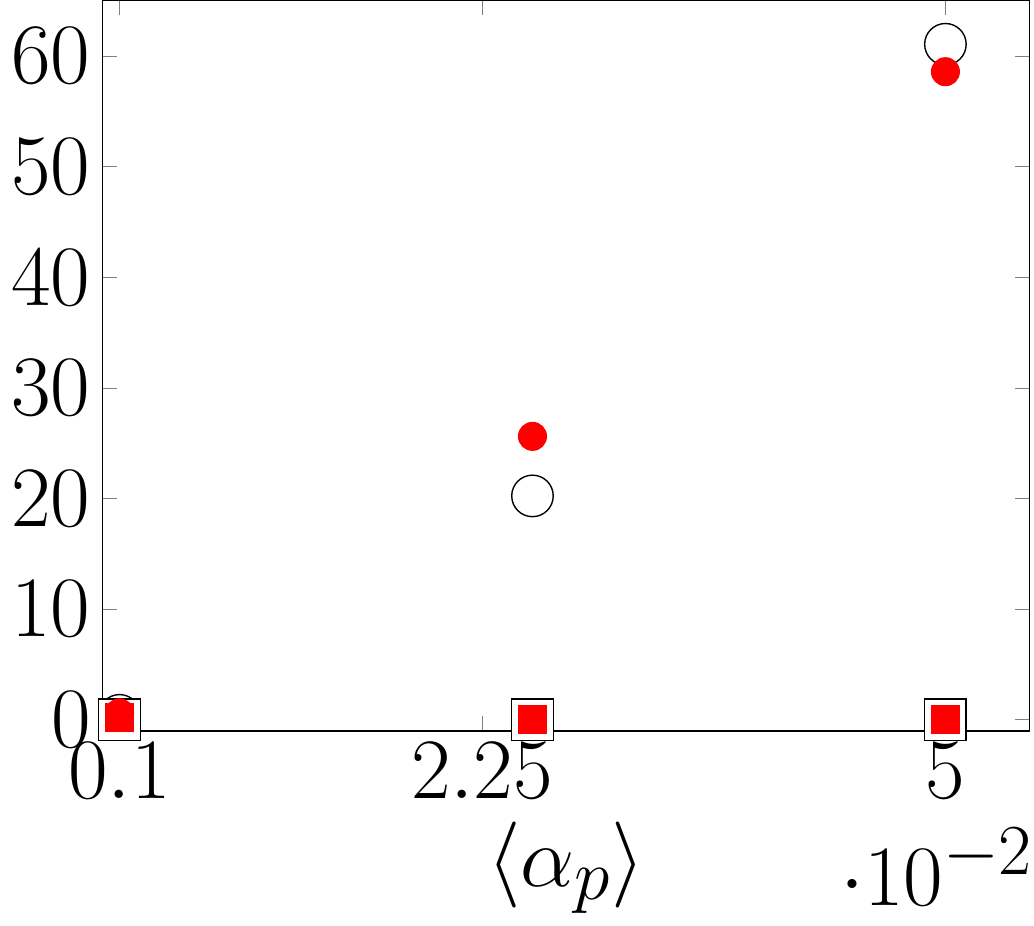}
      }
 \subcaptionbox{$\Arc = 18.0$\label{DP3Simpler}}
     { \includegraphics[height = 0.24\textwidth]{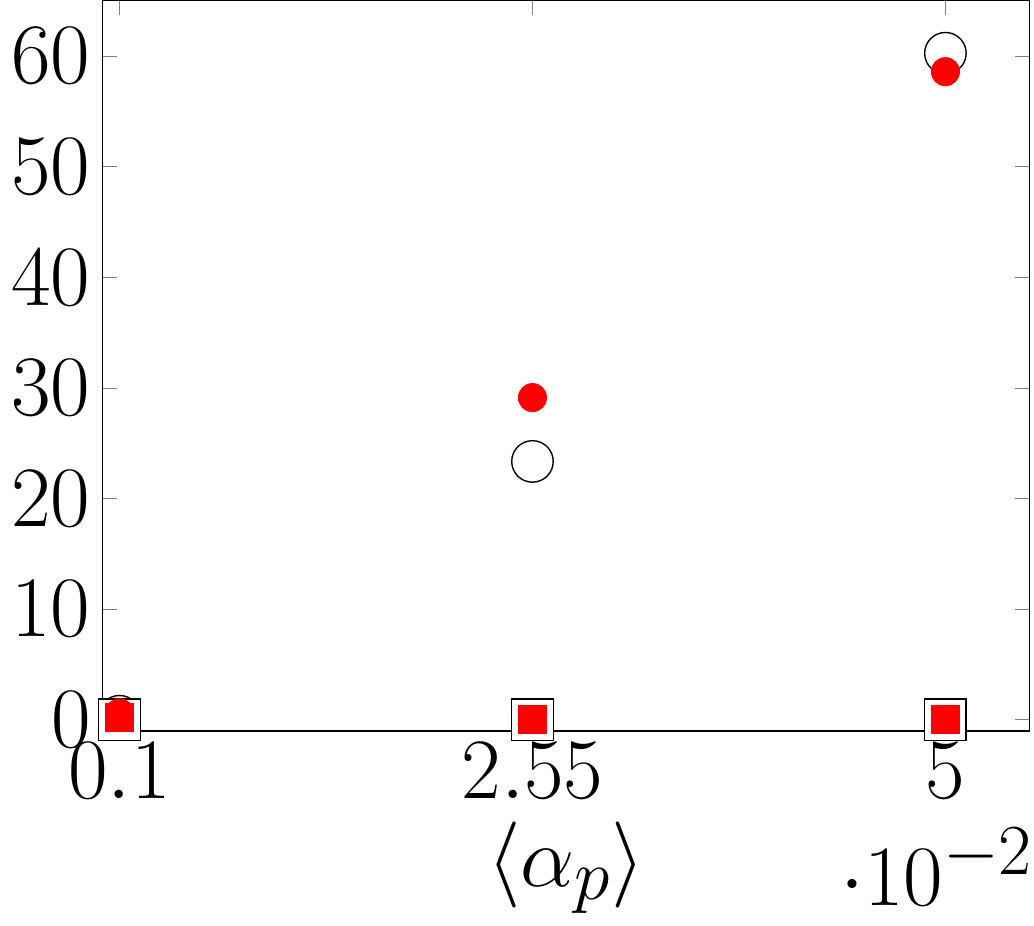}
      } \\
      \vspace{2em}
Six-term model \\
\vspace{1em}
 \subcaptionbox{$\Arc = 1.8$\label{DP1}}
     { \includegraphics[height = 0.24\textwidth]{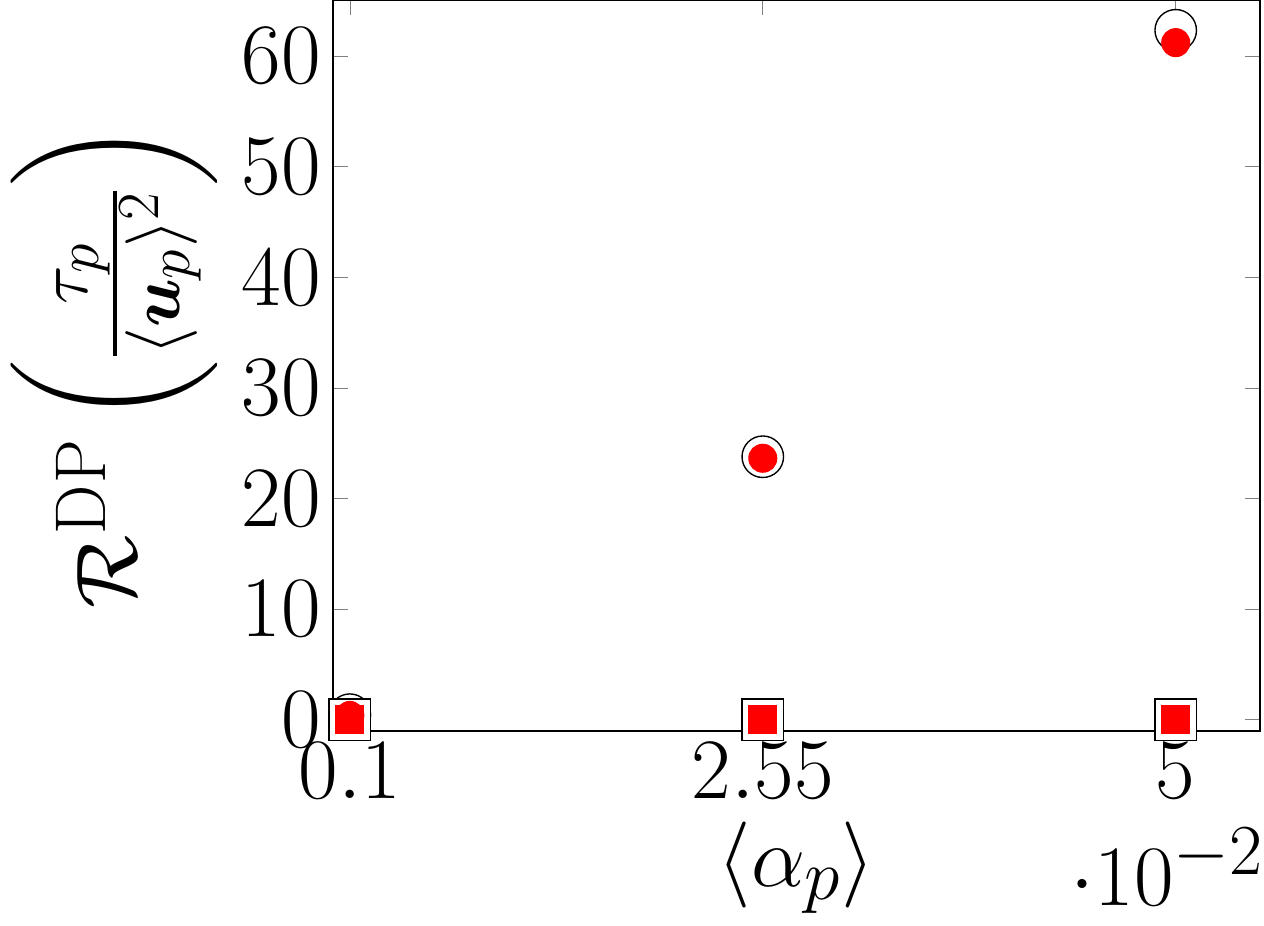}
      }
 \subcaptionbox{$\Arc = 5.4$\label{DP2}}
     { \includegraphics[height = 0.24\textwidth]{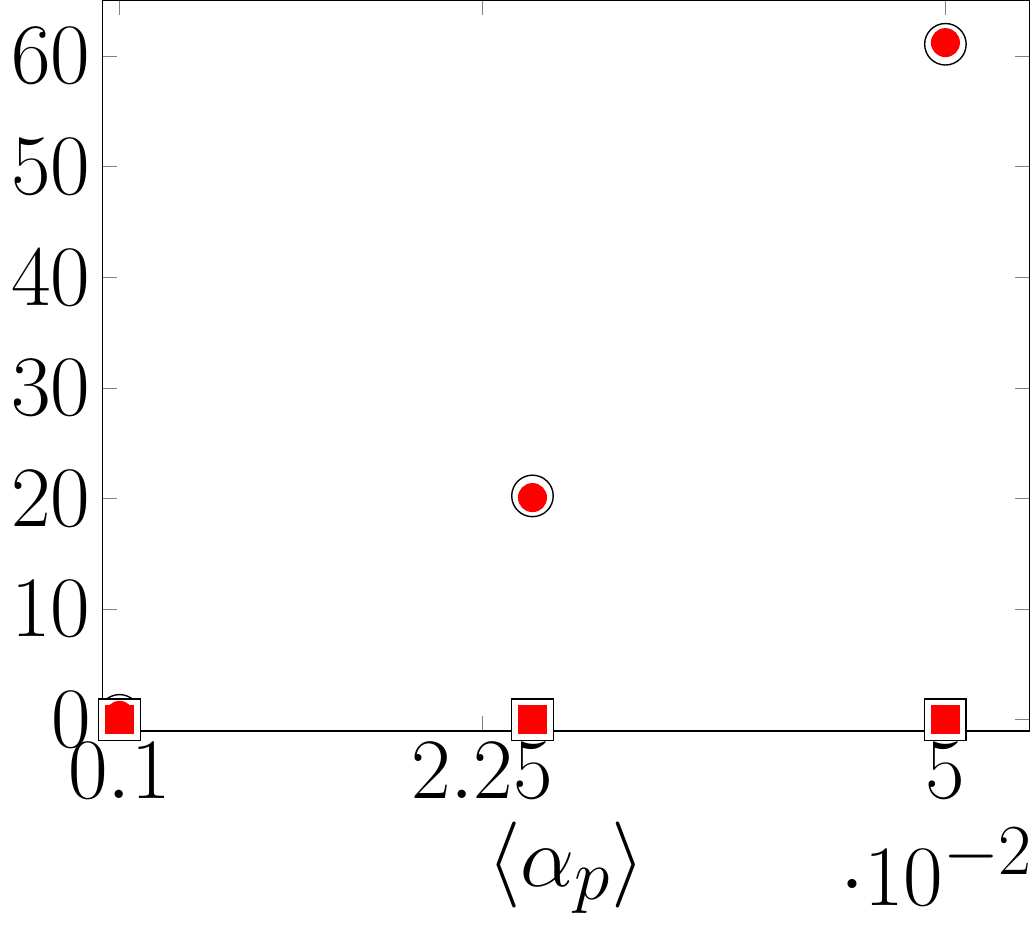}
      }
 \subcaptionbox{$\Arc = 18.0$\label{DP3}}
     { \includegraphics[height = 0.24\textwidth]{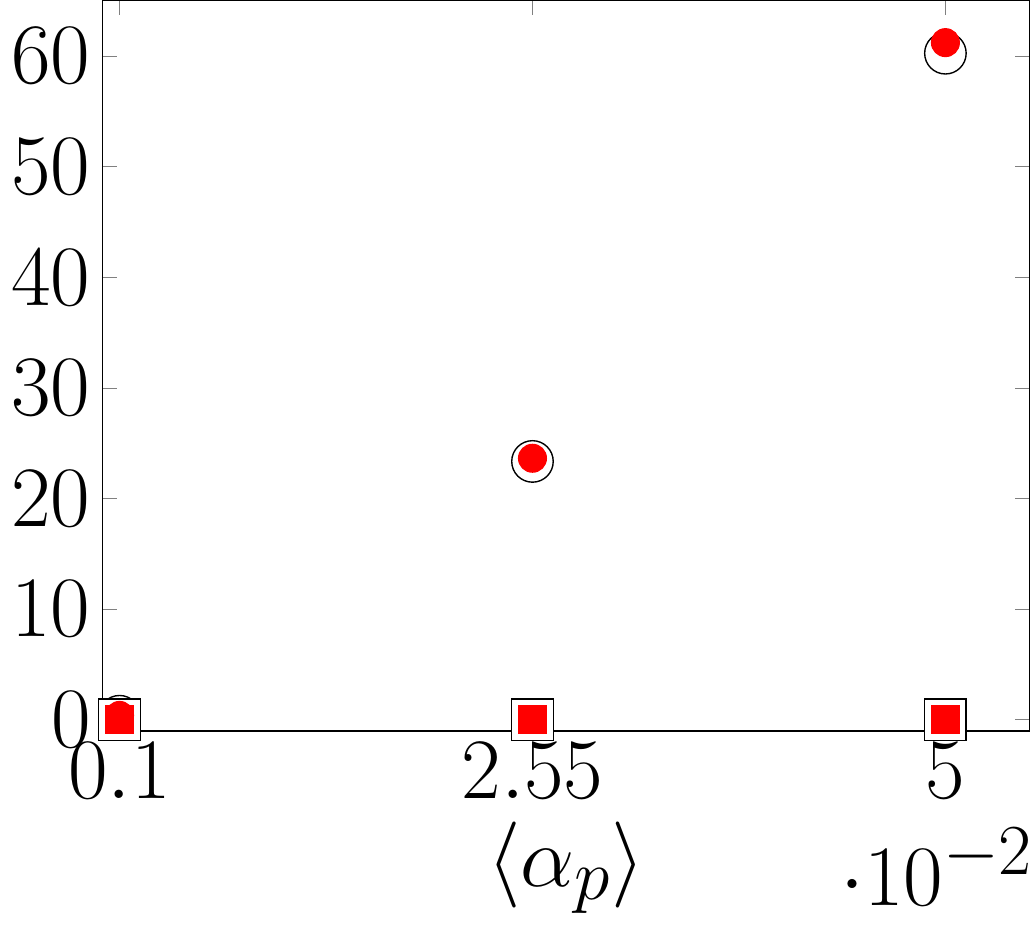}
      }
\caption{Drag production obtained from Eulerian--Lagrangian results (\protect $\square$, cross-stream component and \protect $\circ$, streamwise components) and model prediction (\protect \LearnedSquare, cross-stream component and \protect \LearnedCirc, streamwise components). The model corresponds to Eq.~\ref{eq:DPmodel} with $\lambda = 0.01$. The associated model error is $\epsilon=0.01$.}
\label{fig:DP}
\end{figure}

\begin{figure} 
\centering 
\subcaptionbox{streamwise three-term model \label{DP3term11}}
     { 
    \includegraphics[width = 0.45\textwidth]{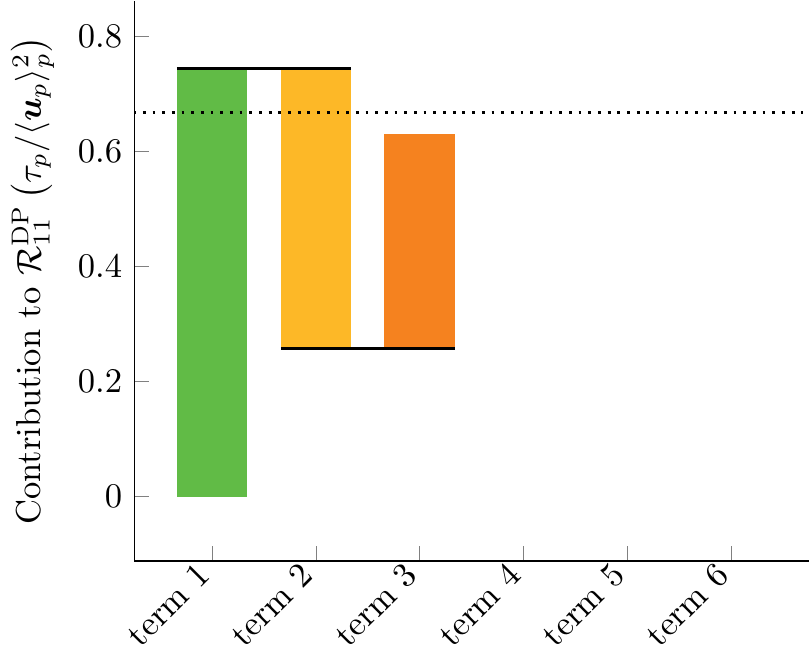}
} \hspace{1em}
\subcaptionbox{streamwise six-term model \label{DP6term11}}
     { 
    \includegraphics[width = 0.45\textwidth]{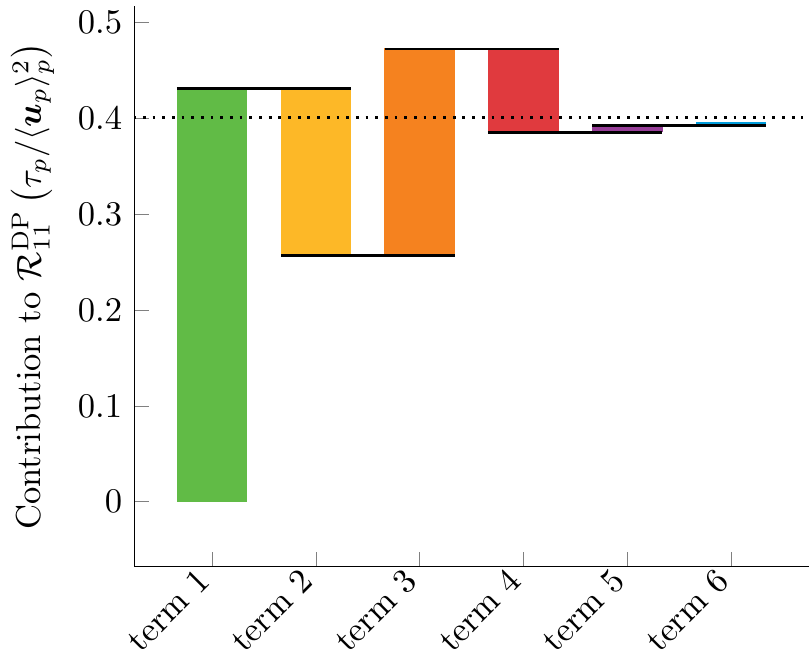}
} \\
\subcaptionbox{Cross-stream three-term model \label{DP3term22}}
     { 
    \includegraphics[width = 0.45\textwidth]{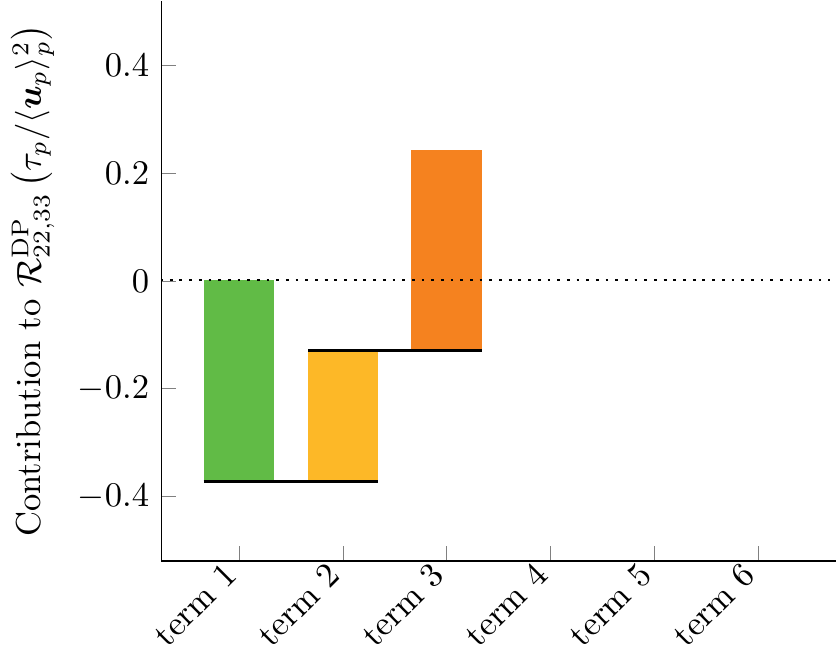}
} \hspace{1em}
\subcaptionbox{Cross-stream six-term model \label{DP6term22}}
     { 
    \includegraphics[width = 0.45\textwidth]{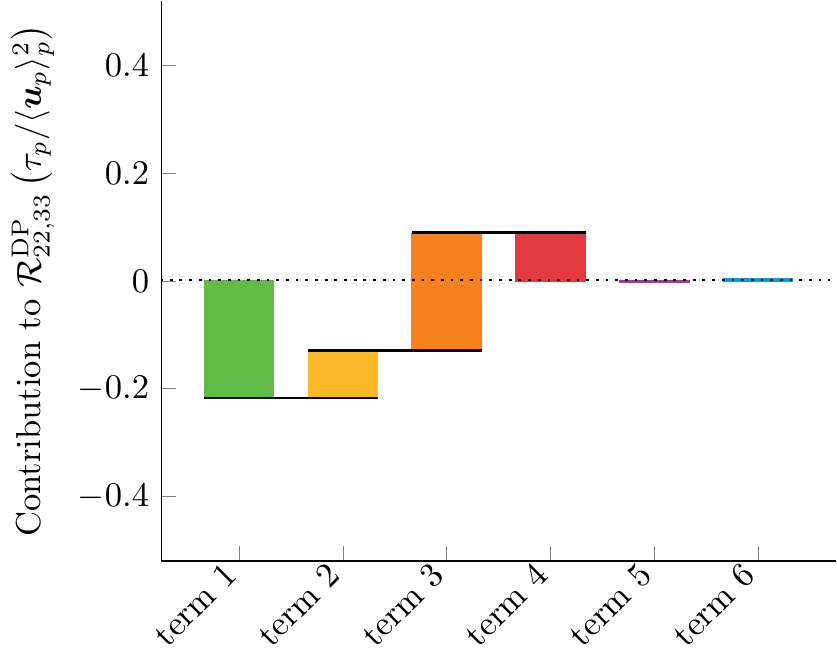}
} 
\caption{Term contributions for the streamwise component of drag production for the three-term (Eq.~\ref{eq:DPmodelSimpler}) and six-term (Eq.~\ref{eq:DPmodel}) models, shown for the case $Ar = 5.40$ and $\langle \alpha_p \rangle = 0.001$. Drag production obtained from the Eulerian--Lagrangian simulations is shown as the dotted line. Terms 1--6 are represented as \protect \TermOneBlock, \protect \TermTwoBlock, \protect \TermThreeBlock, \protect \TermFourBlock, \protect \TermFiveBlock \; and \protect \TermSixBlock, respectively.}
\label{fig:DPcostwalk}
\end{figure}

The resultant learned models with three terms and six terms are given, respectively, as 
\begin{align}
\mathcal{R}^{\text{DP}} &= \frac{\langle u_p \rangle_p^2}{\tau_p} \left \lbrack  \underbrace{1.11 \varphi \a}_{\text{term 1}}  - \underbrace{0.73 \varphi^{-2} \a}_{\text{term 2}} + \underbrace{0.37 \varphi \mathbb{I}}_{\text{term 3}} \right \rbrack \label{eq:DPmodelSimpler} \\ \intertext{and}
\mathcal{R}^{\text{DP}} &= \frac{\langle u_p \rangle_p^2}{\tau_p} \left \lbrack  \underbrace{0.65 \varphi \a}_{\text{term 1}}  - \underbrace{0.26 \varphi^{-2} \a}_{\text{term 2}} + \underbrace{0.22 \varphi \mathbb{I}}_{\text{term 3}} - \underbrace{0.09 \varphi^{-2} \mathbb{I}}_{\text{term 4}} + \underbrace{0.01 \varphi^2 \a}_{\text{term 5}} + \underbrace{0.003 \varphi^2 \mathbb{I}}_{\text{term 6}} \right \rbrack . \label{eq:DPmodel}
\end{align}
To illustrate the interplay between model complexity and interaction, we consider the highly accurate, six-term model (see figures \ref{DP1}--\ref{DP3} and Eq.~\ref{eq:DPmodel}) and the simpler, three-term model (see figures \ref{DP1Simpler}--\ref{DP3Simpler} and Eq.~\ref{eq:DPmodelSimpler}). In comparing the performance of these models, we observe that the general scaling and spread of the data is captured reasonably well with the three-term model, but that the complexity added in the six-term model makes smaller adjustments that drive down model error. As shown in figure~\ref{fig:DPcostwalk}, the accuracy of the three-term model is primarily centered on the streamwise component of drag production (see figure~\ref{DP3term11}); however, it over predicts the cross-stream components (see figure~\ref{DP3term22}). The six-term model, in turn, reduces overall model error by more accurately describing both components; however, this is most pronounced in the cross-stream direction (see figures \ref{DP6term11} and \ref{DP6term22}). 

\begin{figure}
\centering
 \subcaptionbox{$\Arc = 1.8$\label{DP1_train3}}
     { \includegraphics[height = 0.24\textwidth]{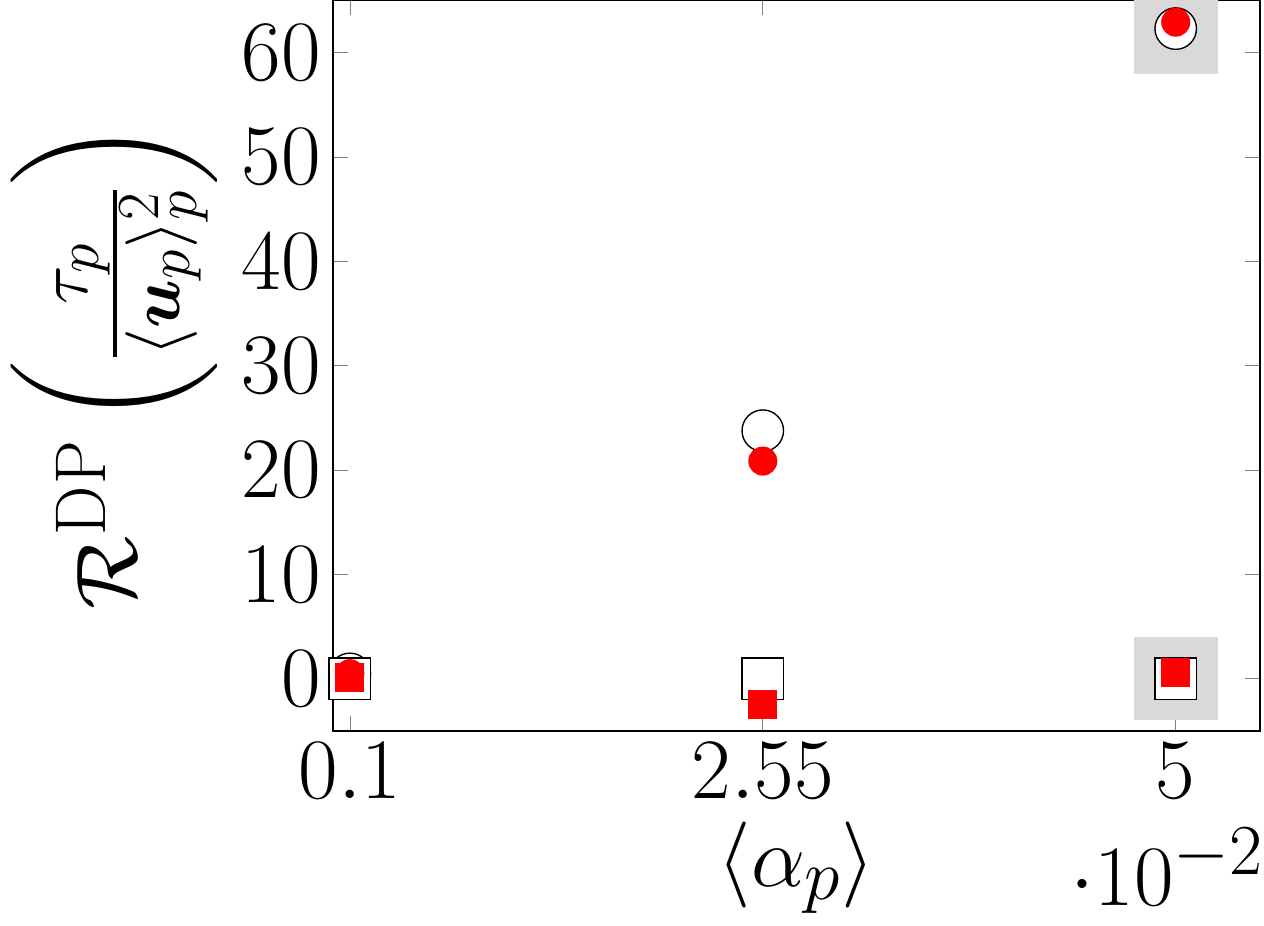}
      }
 \subcaptionbox{$\Arc = 5.4$\label{DP2_train3}}
     { \includegraphics[height = 0.24\textwidth]{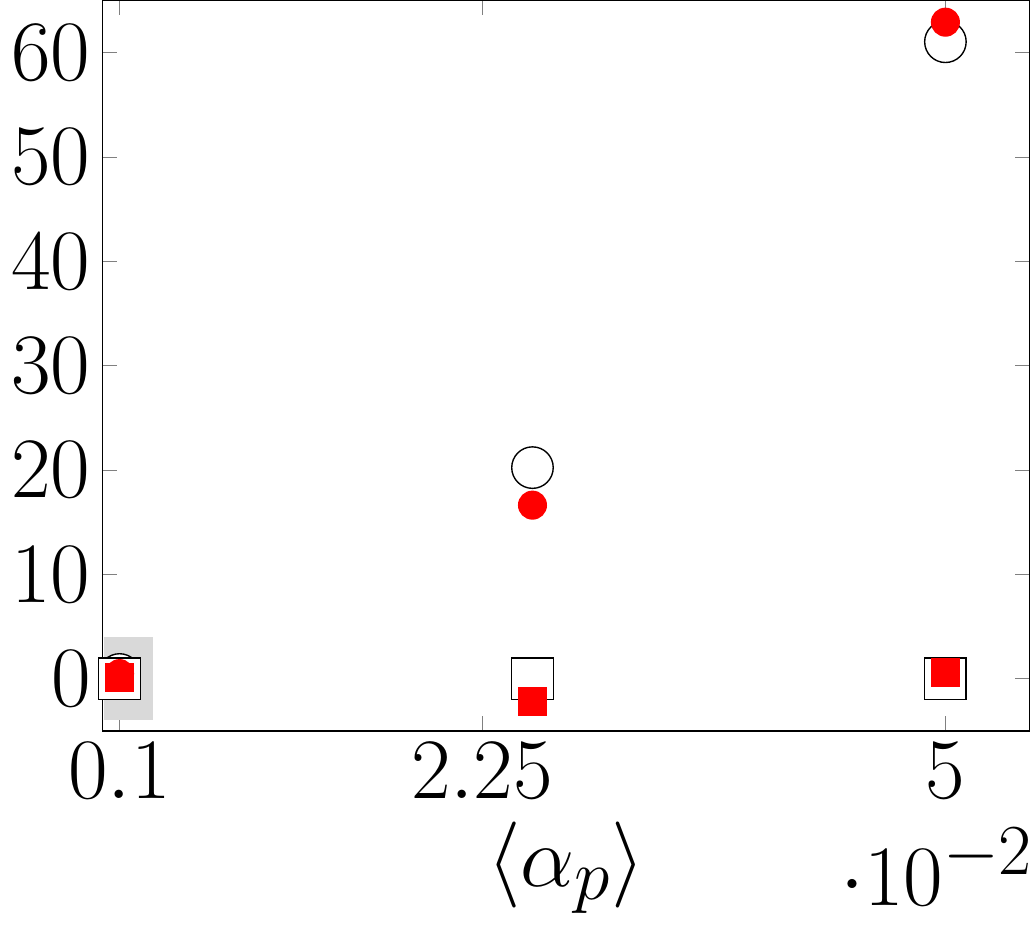}
      }
 \subcaptionbox{$\Arc = 18.0$\label{DP3_train3}}
     { \includegraphics[height = 0.24\textwidth]{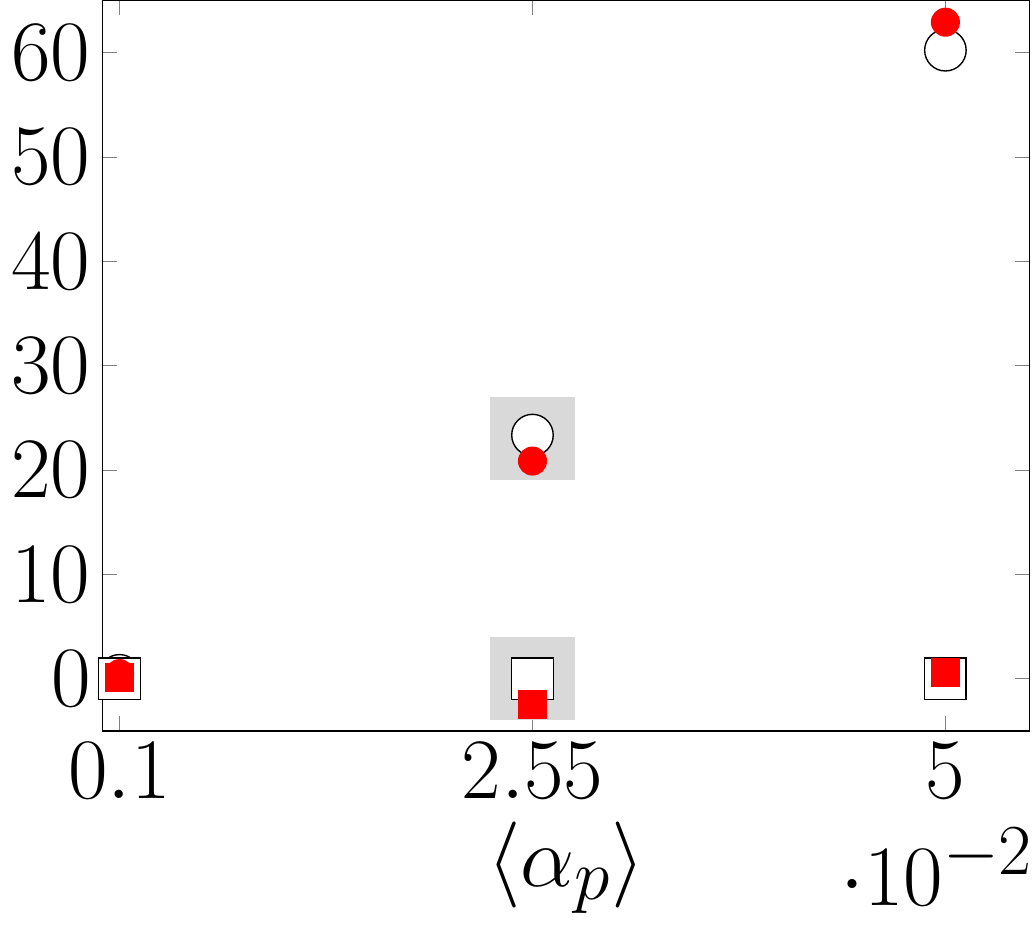}
      }

\caption{Model learned from sparse training data (denoted with grey shaded boxes). The training and testing error are 0.07 and 0.08, respectively. Using the convention from previous figures, Eulerian--Lagrangian results (\protect $\circ$, streamwise component and \protect $\square$, cross-stream components) and model prediction (\protect \LearnedCirc, streamwise component and \protect \LearnedSquare, cross-stream components). The sparsely trained model corresponds to Eq.~\ref{eq:SparseTrainedDP}.}
\label{fig:SparseTrain}
\end{figure}

In addition to discovering compact, algebraic models, sparse regression is also robust to sparse training data \citep{Beetham2020}. To illustrate this, a model was discovered using a sparse training dataset corresponding to $(\Arc, \langle \alpha_p \rangle) = \lbrack (1.8, 0.05), (5.4, 0.001), (18.0, 2.55) \rbrack$ and then tested using the remaining six cases. The resultant model is given as
\begin{equation}
\mathcal{R}^{\text{DP}} = \frac{\langle \bm{u}_p \rangle^2}{\tau_p}  \left \lbrack  \left(0.36 \varphi^{-1} + 0.05 \varphi^2 - 4\times10^{-4} \varphi^3 \right)\a + \left(0.01\varphi^2 + 0.21\varphi^{-1}\right) \mathbb{I} \right \rbrack \label{eq:SparseTrainedDP}
\end{equation}
and shown compared with the trusted Eulerian--Lagrangian data in figures \ref{DP1_train3}--\ref{DP3_train3}. It is notable that the sparsely trained model achieves reasonable accuracy outside the scope of its training and suggests that learned models may be useful even outside the extent of their training, a principal challenge for any model. 

The remaining terms, pressure strain, viscous dissipation and drag exchange, exhibit similar performance as drag production and are summarized here. All three terms are normalized by $k_f/\tau_p$ in order to ensure realizability in the Reynolds stresses. The coefficients supplied to the algorithm as potential contributions to the ideal coefficients, $\hat{\beta}$ included constant coefficients, polynomials in each of $\mathcal{S}^{(i)}$ from -3 to 3 and combinations thereof. 

Pressure strain and viscous diffusion both redistribute turbulent kinetic energy throughout the fluid phase and are present in single-phase flows. The learned models are given as 
\begin{equation}
\mathcal{R}^{\text{PS}} = \varphi \frac{k_f}{\tau_p} \left \lbrack  \langle \alpha_p \rangle \left( 14.36 \a - 22.65 \c \right) +  \langle \alpha_f \rangle \left( 2.60 \c - 2.72 \b \right) \right \rbrack \label{eq:PSmodel}
\end{equation} 
and
\begin{align}\label{eq:VDmodel}
\mathcal{R}^{\text{VD}} = \frac{k_f}{\tau_p}  \Big[ &-1.62 \b\c\c +  \varphi \langle \alpha_p \rangle \left( 0.53\mathbb{I} + 0.72 \a - 3.14 \b\c \right)   \\ \nonumber 
&+\varphi \langle \alpha_f \rangle \left( 0.74 \b - 0.62 \a \right) \c \Big],
\end{align}
respectively. The dominant terms important for capturing the behaviour of pressure strain across flow parameters are $\varphi \langle \alpha_p \rangle \a$ and $\varphi \langle \alpha_p \rangle \c$ and inclusion of only these two terms results in model error of $\epsilon=0.15$. 

\begin{figure}
\centering
 \subcaptionbox{$\Arc = 1.8$\label{PS1}}
     { \includegraphics[height = 0.24\textwidth]{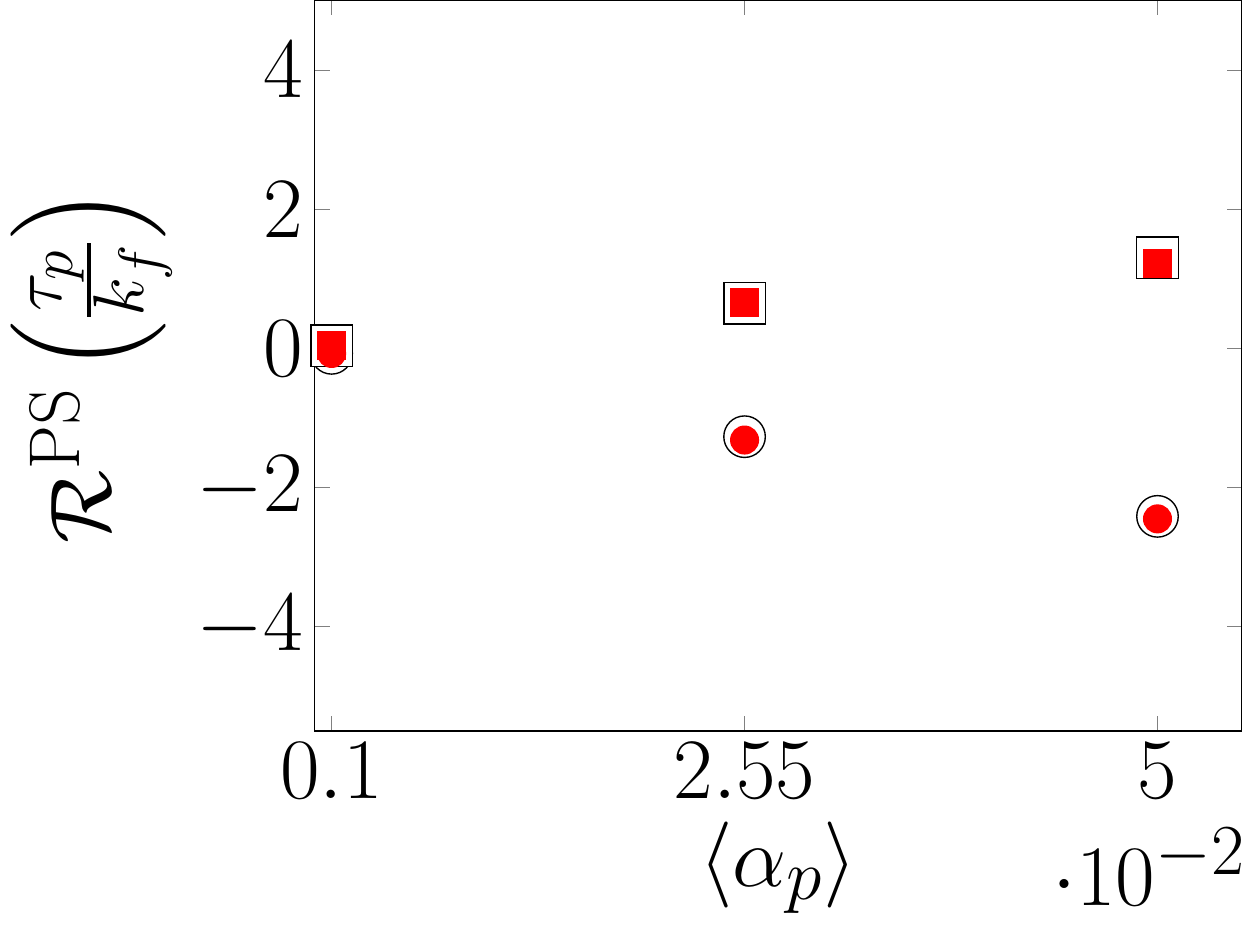}
      }
 \subcaptionbox{$\Arc = 5.4$\label{PS2}}
     { \includegraphics[height = 0.24\textwidth]{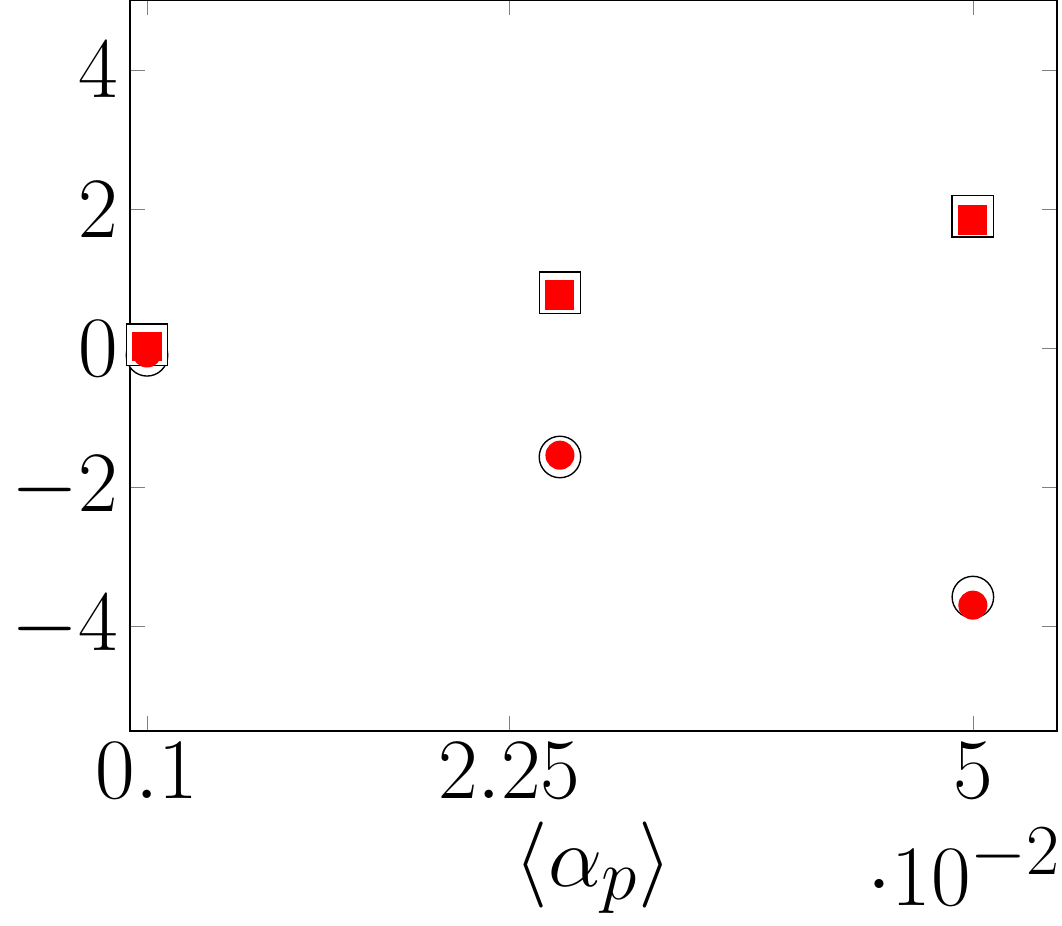}
      }
 \subcaptionbox{$\Arc = 18.0$\label{PS3}}
     { \includegraphics[height = 0.24\textwidth]{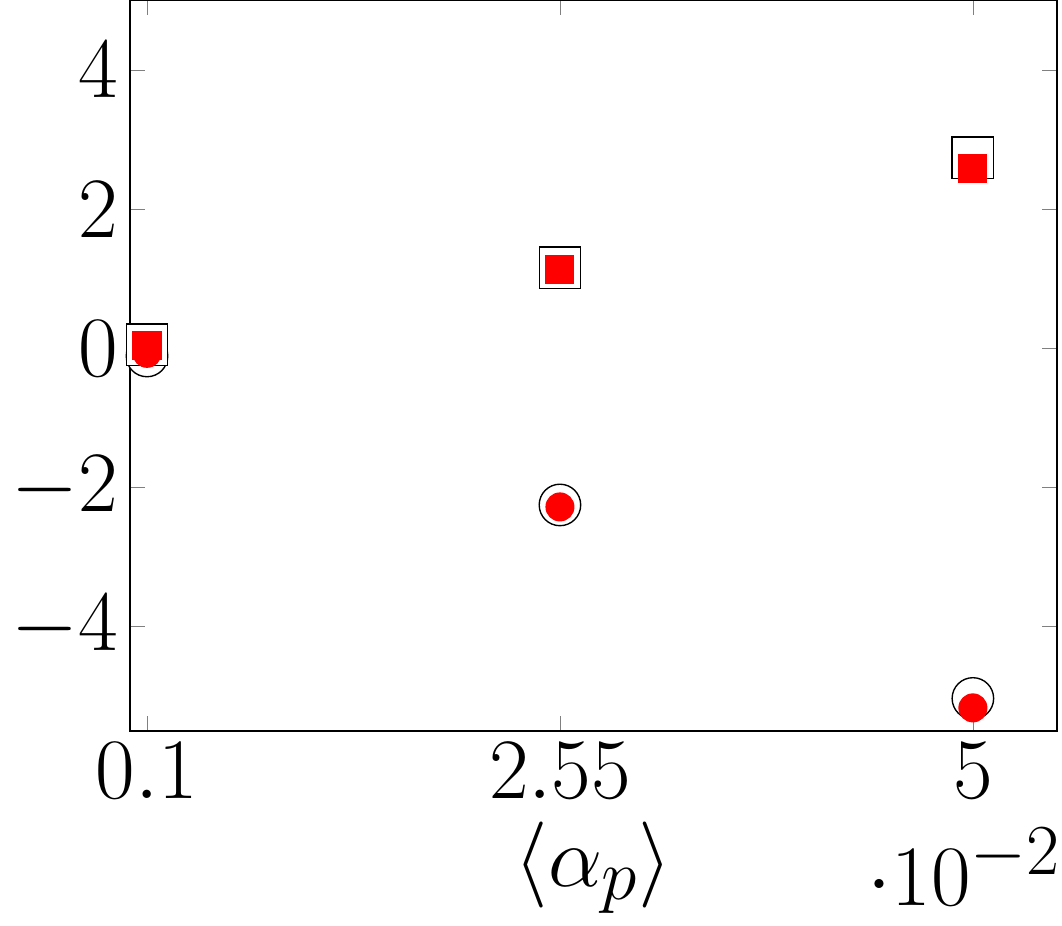}
      }
\caption{Pressure strain Eulerian--Lagrangian results (\protect $\circ$, streamwise component and \protect $\square$, cross-stream components) and model prediction (\protect \LearnedCirc, streamwise component and \protect \LearnedSquare, cross-stream components). Model corresponds to Eq.~\ref{eq:PSmodel} and results from $\lambda = 0.3$. The associated model error is 0.04.}
\label{fig:PS}
\end{figure}

In the case of viscous diffusion, a four-term model is learned in which the three terms that persist into the six-term model are $\varphi \langle \alpha_p \rangle \a$, $\varphi \langle \alpha_p \rangle \b\c$ and $\b\c\c$. The fourth term, $\varphi \langle \alpha_p \rangle \c$, is replaced by the three remaining terms that appear in Eq.~\ref{eq:VDmodel}. This reduces model error from $0.29$ to $0.07$, in a similar manner as described for drag production. 

\begin{figure}
\centering
 \subcaptionbox{$\Arc = 1.8$\label{VD1}}
     { \includegraphics[height = 0.24\textwidth]{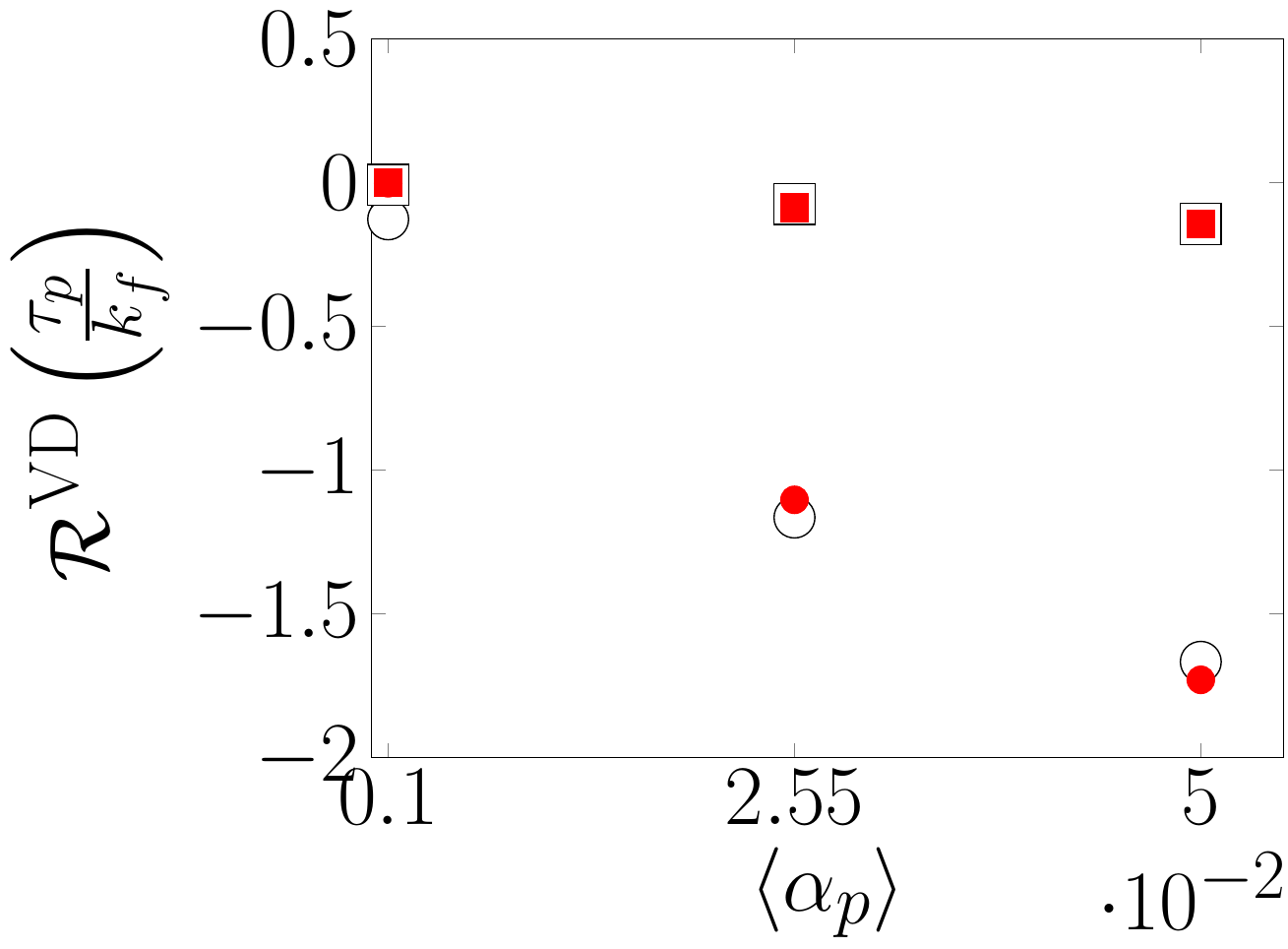}
      }
 \subcaptionbox{$\Arc = 5.4$\label{VD2}}
     { \includegraphics[height = 0.24\textwidth]{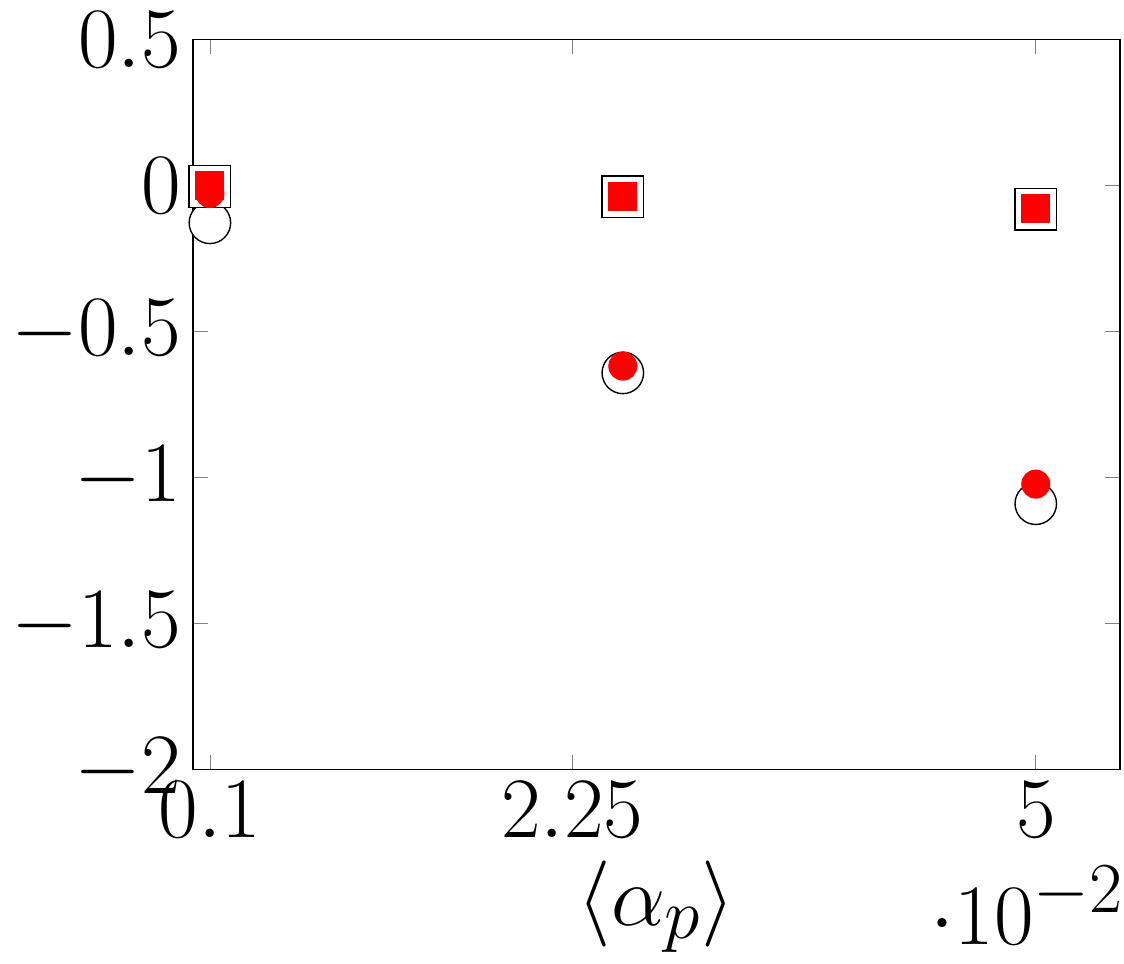}
      }
 \subcaptionbox{$\Arc = 18.0$\label{VD3}}
     { \includegraphics[height = 0.24\textwidth]{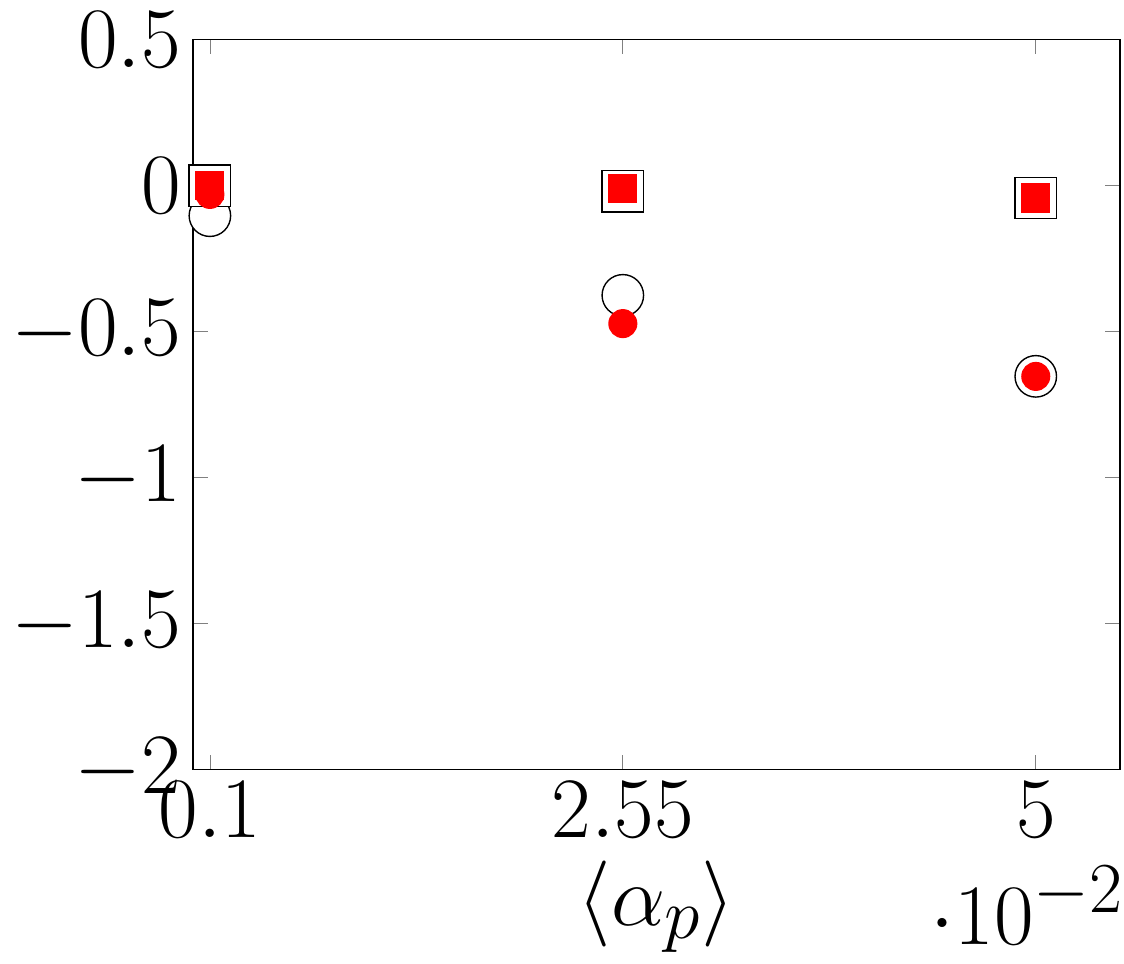}
      }
\caption{Viscous diffusion Eulerian--Lagrangian results (\protect $\circ$, streamwise component and \protect $\square$, cross-stream components) and model prediction (\protect \LearnedCirc, streamwise component and \protect \LearnedSquare, cross-stream components). Model corresponds to Eq.~\ref{eq:VDmodel} and results from $\lambda = 0.2$. The associated model error is 0.07.}
\label{fig:VD}
\end{figure}


Drag exchange describes the mechanism by which turbulent kinetic energy is partitioned between the phases. For this case, all terms in the model are of nearly equal importance. A four-term model is learned which excludes $\Arc\a$ with an error of $\epsilon=0.16$ as compared with the model error of $\epsilon=0.15$ in the case of the five-term model. This is due to the minimal dependence of the data on Archimedes number. 

\begin{figure}
\centering
 \subcaptionbox{$\Arc = 1.8$\label{DE1}}
     { \includegraphics[height = 0.24\textwidth]{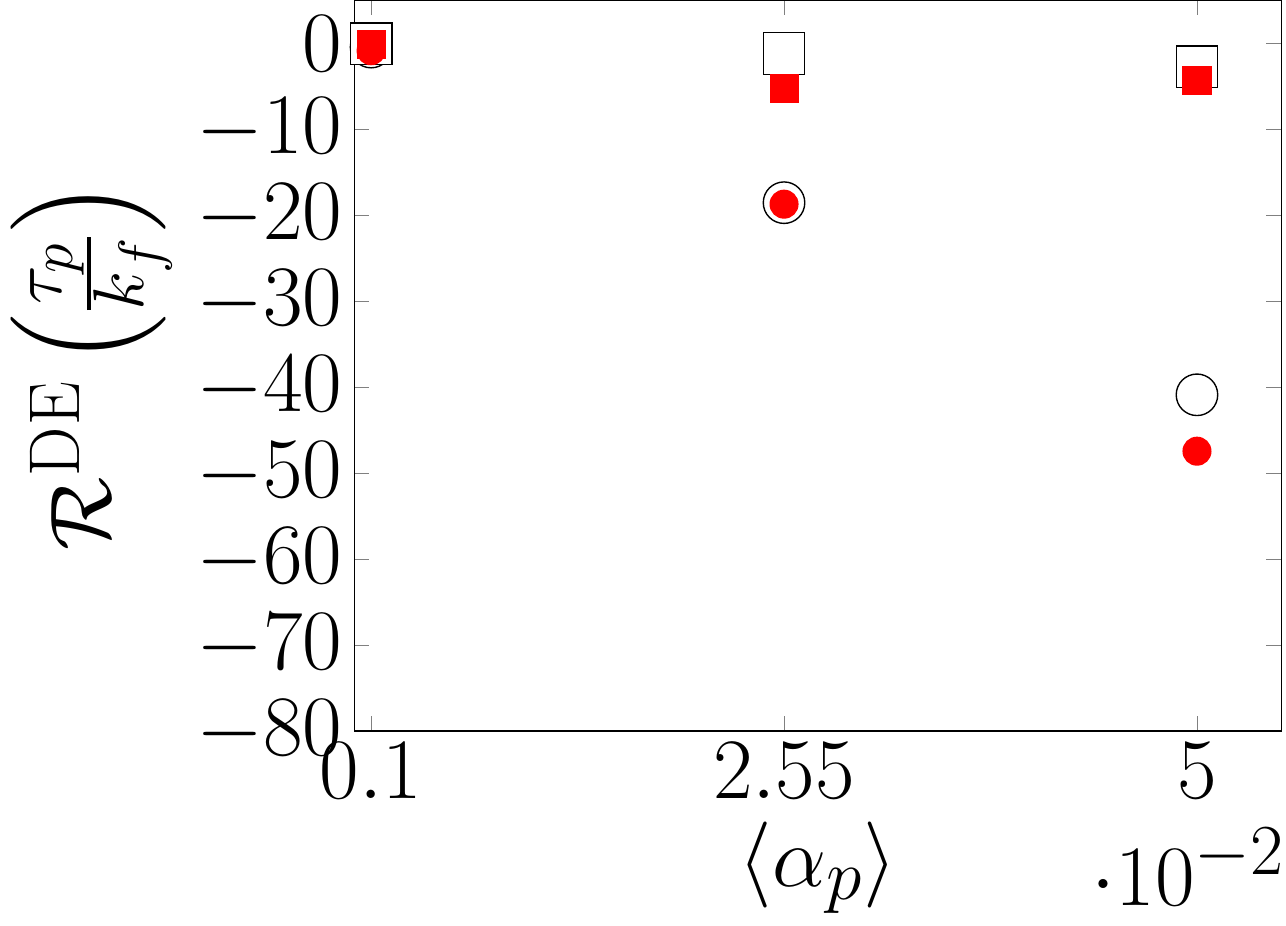}
      }
 \subcaptionbox{$\Arc = 5.4$\label{DE2}}
     { \includegraphics[height = 0.24\textwidth]{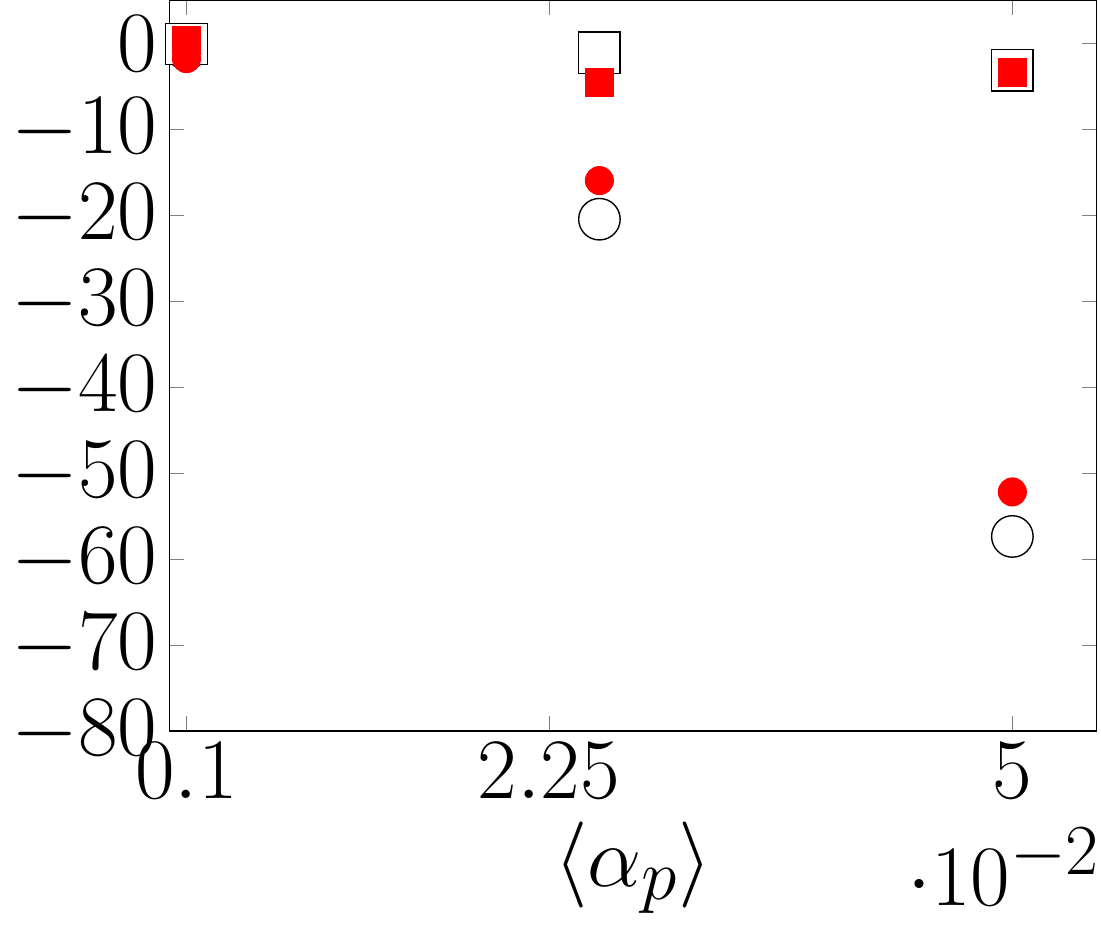}
      }
 \subcaptionbox{$\Arc = 18.0$\label{DE3}}
     { \includegraphics[height = 0.24\textwidth]{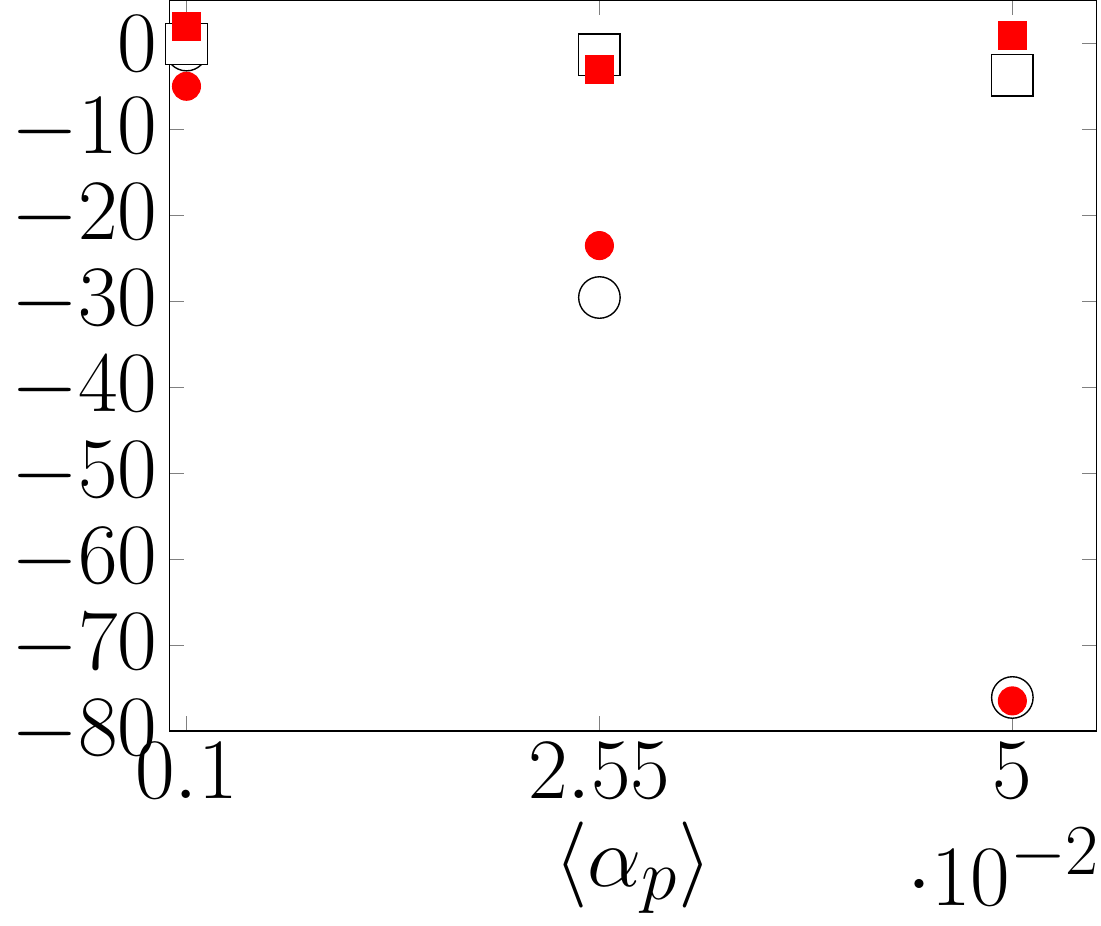}
      }
\caption{Drag exchange Eulerian--Lagrangian results (\protect $\circ$, streamwise component and \protect $\square$, cross-stream components) and model prediction (\protect \LearnedCirc, streamwise component and \protect \LearnedSquare, cross-stream components). Model corresponds to Eq.~\ref{eq:DEmodel} and results from $\lambda = 0.006$. The associated model error is 0.15.}
\label{fig:DE}
\end{figure}


\begin{equation}
\mathcal{R}^{\text{DE}} =  \varphi \frac{k_f}{\tau_p}  \left \lbrack  -0.36 \varphi \mathbb{I} - \left(0.02 \varphi^{2} \langle \alpha_f \rangle^2 +0.38 \Arc \right)\a + 0.04 \varphi^3 \langle \alpha_f \rangle^3 \b \left(\b - \c \right) \c \right \rbrack \label{eq:DEmodel} 
\end{equation}

For all of the terms considered, sparse regression is capable of uncovering models with model error to machine precision of zero (associated with $\lambda = 0$); however, these resultant models are substantially more complex and likely would not perform well outside the scope of training due to overfitting subtle nonlinearities. These models, for comparison, contain 18 terms for pressure strain, viscous diffusion and drag exchange, respectively, and 8 terms for drag production.

\section{Conclusions} 
In this work, the multiphase RANS equations are presented for two-way coupled gas--solid flows. In this class of flows, the coupling between the phases spontaneously gives rise to coherent particle structures, which in turn generate and sustain turbulence in the carrier phase. This phenomenon has important engineering implications \citep{shaffer2013high,miller2014carbon,Guo2019,Beetham2019} and makes the formulation of closure models that are predictive across scales and flow conditions challenging.

We apply a newly formulated modelling methodology, sparse regression with embedded form invariance \citep{Beetham2020}, to highly-resolved Eulerian--Lagrangian data for fully-developed CIT. The benefits of this methodology as compared with Neural Networks, which have become increasingly popular, are (1) interpretability of the resultant closures, since they are in a closed algebraic formulation, (2) ease of dissemination to existing RANS solvers and (3) robustness to very sparse training sets. The dataset used for model development spans a range of flow parameters, specifically $\Arc = (1.8, 5.4, 18.0)$ and $\langle \alpha_p \rangle = (0.001, 0.0255, 0.05)$, in order to formulate models across a range of typical conditions. 

Particular attention is paid to the closures for the four dominant unclosed terms that appear in the fluid-phase Reynolds stress equations -- pressure strain, viscous diffusion, drag production and drag exchange. In applying the sparse regression method to each of these terms individually, we discover compact closures containing between four and six term that are accurate across the scope of training (model error ranges from 0.01 to 0.15). Because of the compact nature of the models developed and the nature of the sparse regression algorithm, we are able to assess the relative importance of each term and its role in reducing model error. Further, we demonstrate that even when training on a subset of the Eulerian--Lagrangian data, the methodology learns a model that remains accurate outside the scope of its training. Additionally, because of the compact, algebraic formulation of the method, resultant models are accessible for interpretation and terms of greater physical significance are easily identified.  These findings suggest that the sparse regression methodology holds promise for developing closures for more complicated multiphase flows, such as channel, duct or bubbly flows. These will be the subject of future investigation.

\bibliographystyle{jfm}
\bibliography{SparseRegression_CIT}

\begin{thebibliography}{47}
\expandafter\ifx\csname natexlab\endcsname\relax\def\natexlab#1{#1}\fi
\def\au#1{#1} \def\ed#1{#1} \def\yr#1{#1}\def\at#1{#1}\def\jt#1{\textit{#1}}
  \def\bt#1{#1}\def\bvol#1{\textbf{#1}} \def\vol#1{#1} \def\pg#1{#1}
  \def\publ#1{#1}\def\arxiv#1{#1}\def\org#1{#1}\def\st#1{\textit{#1}}

\bibitem[Anderson \& Jackson(1967)]{Anderson1967}
{\sc \au{Anderson, T.~B.} \& \au{Jackson, R.}} \yr{1967}  \at{Fluid mechanical
  description of fluidized beds. {E}quations of motion}.  \jt{Industrial \&
  Engineering Chemistry Fundamentals}  \bvol{6}~(4),  \pg{527--539}.

\bibitem[Batchelor(1972)]{Batchelor1972}
{\sc \au{Batchelor, G.~K.}} \yr{1972}  \at{Sedimentation in a dilute dispersion
  of spheres}.  \jt{Journal of Fluid Mechanics}  \bvol{52},  \pg{245--268}.

\bibitem[Batchelor(1982)]{Batchelor1982}
{\sc \au{Batchelor, G.~K.}} \yr{1982}  \at{Sedimentation in a dilute
  polydisperse system of interacting spheres. {P}art 1. {G}eneral theory}.
  \jt{Journal of Fluid Mechanics}  \bvol{119},  \pg{379--408}.

\bibitem[Batchelor(1988)]{Batchelor1988}
{\sc \au{Batchelor, G.~K.}} \yr{1988}  \at{A new theory of the instability of a
  uniform fluidized bed}.  \jt{Journal of Fluid Mechanics}  \bvol{193},
  \pg{75--110}.

\bibitem[Beetham \& Capecelatro(2019)]{Beetham2019}
{\sc \au{Beetham, S.} \& \au{Capecelatro, J.}} \yr{2019}  \at{Biomass pyrolysis
  in fully-developed turbulent riser flow}.  \jt{Renewable Energy}  \bvol{140},
   \pg{751--760}.

\bibitem[Beetham \& Capecelatro(2020)]{Beetham2020}
{\sc \au{Beetham, S.} \& \au{Capecelatro, J.}} \yr{2020}  \at{Formulating
  turbulence closures using sparse regression with embedded form invariance} ,
  \arxiv{arXiv: 2003.12884}.

\bibitem[Bode {\em et~al.\/}(2019)Bode, Gauding, Kleinheinz \&
  Pitsch]{Bode2019}
{\sc \au{Bode, M.}, \au{Gauding, M.}, \au{Kleinheinz, K.} \& \au{Pitsch, H.}}
  \yr{2019}  \at{Deep learning at scale for subgrid modeling in turbulent
  flows: regression and reconstruction}.  \jt{arXiv:1910.00928v1} .

\bibitem[Brunton {\em et~al.\/}(2016)Brunton, Proctor \& Kutz]{ML_2016Brunton}
{\sc \au{Brunton, S.~L.}, \au{Proctor, J.~L.} \& \au{Kutz, J.~N.}} \yr{2016}
  \at{Discovering governing equations from data by sparse identification of
  nonlinear dynamical systems.}  \jt{Proceedings of the National Academy of
  Sciences}  \bvol{113(15)},  \pg{3932--3937}.

\bibitem[Cao \& Ahmadi(1995)]{cao1995}
{\sc \au{Cao, J.} \& \au{Ahmadi, G.}} \yr{1995}  \at{Gas--particle two--phase
  turbulent flow in a vertical duct}.  \jt{International Journal of Multiphase
  Flow}  \bvol{21}~(6),  \pg{1203--1228}.

\bibitem[Capecelatro \& Desjardins(2013)]{Capecelatro2013}
{\sc \au{Capecelatro, J.} \& \au{Desjardins, O.}} \yr{2013}  \at{An
  {E}uler--{L}agrange strategy for simulating particle-laden flows}.
  \jt{Journal of Computational Physics}  \bvol{238},  \pg{1--31}.

\bibitem[Capecelatro {\em et~al.\/}(2006)Capecelatro, Desjardins \&
  Fox]{capecelatro2016channel2}
{\sc \au{Capecelatro, J.}, \au{Desjardins, O.} \& \au{Fox, R.~O.}} \yr{2006}
  \at{Strongly--coupled gas--particle flows in vertical channels. part ii:
  Turbulence modeling}.  \jt{Physics of Fluids}  \bvol{28},  \pg{1--22}.

\bibitem[Capecelatro {\em et~al.\/}(2015)Capecelatro, Desjardins \&
  Fox]{capecelatro2015}
{\sc \au{Capecelatro, J.}, \au{Desjardins, O.} \& \au{Fox, R.~O.}} \yr{2015}
  \at{On fluid-particle dynamics in fully-developed cluster-induced
  turbulence}.  \jt{Journal of Fluid Mechanics}  \bvol{780},  \pg{578--635}.

\bibitem[Capecelatro {\em et~al.\/}(2016)Capecelatro, Desjardins \&
  Fox]{Capecelatro2016_domain}
{\sc \au{Capecelatro, J.}, \au{Desjardins, O.} \& \au{Fox, R.~O.}} \yr{2016}
  \at{Effect of domain size on fluid-particle statistics in homogeneous,
  gravity-driven, cluster-induced turbulence}.  \jt{Journal of Fluids
  Engineering}  \bvol{138},  \pg{1--8}.

\bibitem[Cheng {\em et~al.\/}(1999)Cheng, Guo, Wei, Jin \& Lin]{Cheng1999}
{\sc \au{Cheng, Y.}, \au{Guo, Y.}, \au{Wei, F.}, \au{Jin, Y.} \& \au{Lin, W.}}
  \yr{1999}  \at{Modeling the hydrodynamics of downer reactors based on kinetic
  theory}.  \jt{Chemical Engineering Science}  \bvol{54},  \pg{2019--2027}.

\bibitem[Cundall \& Strack(1979)]{Cundall1979}
{\sc \au{Cundall, P.~A.} \& \au{Strack, O. D.~L.}} \yr{1979}  \at{A discrete
  numerical model for granular assemblies}.  \jt{Geotechnique}  \bvol{29}~(1),
  \pg{47--65}.

\bibitem[Dasgupta {\em et~al.\/}(1994)Dasgupta, Jackson \&
  Sundaresan]{dasgupta1994turbulent}
{\sc \au{Dasgupta, S.}, \au{Jackson, R.} \& \au{Sundaresan, S.}} \yr{1994}
  \at{Turbulent gas--particle flow in vertical risers}.  \jt{AIChE Journal}
  \bvol{40}~(2),  \pg{215--228}.

\bibitem[Dasgupta {\em et~al.\/}(1998)Dasgupta, Jackson \&
  Sundaresan]{Dasgupta1998}
{\sc \au{Dasgupta, S.}, \au{Jackson, R.} \& \au{Sundaresan, S.}} \yr{1998}
  \at{Gas--particle flow in vertical pipes with high mass loading of
  particles}.  \jt{Powder Technology}  \bvol{96},  \pg{6--23}.

\bibitem[Desjardins {\em et~al.\/}(2008)Desjardins, Blanquart, Balarac \&
  Pitsch]{desjardins2008high}
{\sc \au{Desjardins, O.}, \au{Blanquart, G.}, \au{Balarac, G.} \& \au{Pitsch,
  H.}} \yr{2008}  \at{High order conservative finite difference scheme for
  variable density low {M}ach number turbulent flows}.  \jt{Journal of
  Computational Physics}  \bvol{227}~(15),  \pg{7125--7159}.

\bibitem[Duraisamy \& Durbin(2014)]{ML_2014Duraisamy_transition}
{\sc \au{Duraisamy, K.} \& \au{Durbin, P.~A.}} \yr{2014}  \at{Transition
  modeling using data driven approaches}.  \jt{Proceedings of the Summer
  Program}  \pg{p. 427}.

\bibitem[Duraisamy {\em et~al.\/}(2015)Duraisamy, Zhang \&
  Singh]{ML_2014Duraisamy_new}
{\sc \au{Duraisamy, K.}, \au{Zhang, Z.~J.} \& \au{Singh, A.~P.}} \yr{2015}
  \at{New approaches in turbulence and transition modeling using data-driven
  techniques}.  \jt{53rd AIAA Aerospace Sciences Meeting}  \pg{p. 1284}.

\bibitem[Fox(2014)]{fox2014}
{\sc \au{Fox, R.~O.}} \yr{2014}  \at{On multiphase turbulence models for
  collisional fluid--particle flows}.  \jt{Journal of Fluid Mechanics}
  \bvol{742},  \pg{368--424}.

\bibitem[Gatski \& Speziale(1993)]{Gatski1993}
{\sc \au{Gatski, T.~B.} \& \au{Speziale, C.~G.}} \yr{1993}  \at{On explicit
  algebraic stress models for complex turbulent flows}.  \jt{Journal of Fluid
  Mechanics}  \bvol{254},  \pg{59--78}.

\bibitem[Guo \& Capecelatro(2019)]{Guo2019}
{\sc \au{Guo, L.} \& \au{Capecelatro, J.}} \yr{2019}  \at{The role of clusters
  on heat transfer in sedimenting gas-solid flows}.  \jt{International Journal
  of Heat and Mass Transfer}  \bvol{132},  \pg{1217--1230}.

\bibitem[Innocenti {\em et~al.\/}(2019)Innocenti, Fox, Salvetti \&
  Chibbaro]{Innocenti2019}
{\sc \au{Innocenti, A.}, \au{Fox, R.~O.}, \au{Salvetti, M.V.} \& \au{Chibbaro,
  S.}} \yr{2019}  \at{A {L}agrangian probability-density-function model for
  collisional turbulent fluid-particle flows}.  \jt{Journal of Fluid Mechanics}
   \bvol{862},  \pg{449--489}.

\bibitem[Jiang \& Zhang(2012)]{Jiang2012}
{\sc \au{Jiang, Y.Y.} \& \au{Zhang, P.}} \yr{2012}  \at{Numerical investigation
  of slush nitrogen flow in a horizontal pipe}.  \jt{Chemical Engineering
  Science}  \bvol{73},  \pg{169--180}.

\bibitem[Jordan \& Mitchell(2015)]{ML_2015Jordan}
{\sc \au{Jordan, M.~I.} \& \au{Mitchell, T.~M.}} \yr{2015}  \at{Machine
  learning: Trends, perspectives, and prospects.}  \jt{Science}
  \bvol{349(6245)},  \pg{255--260}.

\bibitem[Ling {\em et~al.\/}(2016)Ling, Kurzawski \& Templeton]{Ling2016}
{\sc \au{Ling, J.}, \au{Kurzawski, A.} \& \au{Templeton, J.}} \yr{2016}
  \at{Reynolds averaged turbulence modeling using deep neural networks with
  embedded invariance}.  \jt{Journal of Fluid Mechanics}  \bvol{807},
  \pg{155--166}.

\bibitem[Liu \& Fang(2019)]{Liu2019}
{\sc \au{Liu, W.} \& \au{Fang, J.}} \yr{2019}  \at{Iterative framework of
  machine-learning based turbulence modeling for {R}eynolds-averaged
  {N}avier-{S}tokes simulations}.  \jt{arXiv:1910.01232v1} .

\bibitem[Lu(2010)]{ML_2010Lu}
{\sc \au{Lu, C.}} \yr{2010}  \at{Artificial neural network for behavior
  learning from meso-scale simulations, application to multi-scale
  multimaterial flows}.  \jt{PhD thesis} .

\bibitem[Ma {\em et~al.\/}(2016)Ma, Lu \& Tryggvason]{ML_2016Ma}
{\sc \au{Ma, M.}, \au{Lu, J.} \& \au{Tryggvason, G.}} \yr{2016}  \at{Using
  statistical learning to close two-fluid multiphase flow equations for bubbly
  flows in vertical channels}.  \jt{International Journal of Multiphase Flow}
  \bvol{85},  \pg{336--347}.

\bibitem[Milano \& Koumoutsakos(2002)]{ML_2002Milano}
{\sc \au{Milano, M.} \& \au{Koumoutsakos, P.}} \yr{2002}  \at{Neural network
  modeling for near wall turbulent flow}.  \jt{Journal of Computational
  Physics}  \bvol{182},  \pg{1--26}.

\bibitem[Miller {\em et~al.\/}(2014)Miller, Syamlal, Mebane, Storlie,
  Bhattacharyya, Sahinidis, Agarwal, Tong, Zitney, Sarkar, Sun, Sundaresan,
  Ryan, Engel \& Dale]{miller2014carbon}
{\sc \au{Miller, D.~C.}, \au{Syamlal, M.}, \au{Mebane, D.~S.}, \au{Storlie,
  C.}, \au{Bhattacharyya, D.}, \au{Sahinidis, N.~V.}, \au{Agarwal, D.},
  \au{Tong, C.}, \au{Zitney, S.~E.}, \au{Sarkar, A.}, \au{Sun, X.},
  \au{Sundaresan, S.}, \au{Ryan, E.}, \au{Engel, D.} \& \au{Dale, C.}}
  \yr{2014}  \at{Carbon capture simulation initiative: a case study in
  multiscale modeling and new challenges}.  \jt{Annual Review of Chemical and
  Biomolecular Engineering}  \bvol{5},  \pg{301--323}.

\bibitem[Pierce(2001)]{pierce2001progress}
{\sc \au{Pierce, C.~D.}} \yr{2001}  \at{Progress-variable approach for
  large-eddy simulation of turbulent combustion}. PhD thesis, Stanford
  University.

\bibitem[Pope(1975)]{ML_1975Pope}
{\sc \au{Pope, S.~B.}} \yr{1975}  \at{A more general effective-viscosity
  hypothesis}.  \jt{Journal of Fluid Mechanics}  \bvol{72}~(2),  \pg{331--340}.

\bibitem[Pope(2000)]{PopeText}
{\sc \au{Pope, S.~B.}} \yr{2000}  \at{Turbulent flows}.  \jt{Cambridge
  University Press} .

\bibitem[Rajabi \& Kavianpour(2012)]{ML_2012Rajabi}
{\sc \au{Rajabi, E.} \& \au{Kavianpour, M.~R.}} \yr{2012}  \at{Intelligent
  prediction of turbulent flow over backward-facing step using direct numerical
  simulation data}.  \jt{Engineering Applications of Computational Fluid
  Mechanics}  \bvol{6(4)},  \pg{490--503}.

\bibitem[Rao {\em et~al.\/}(2012)Rao, Curtis, Hancock \& Wassgren]{Rao2012}
{\sc \au{Rao, A.}, \au{Curtis, J.~S.}, \au{Hancock, B.~C.} \& \au{Wassgren,
  C.}} \yr{2012}  \at{Numerical simulation of dilute turbulent gas--particle
  flow with turbulence modulation}.  \jt{{AIC}h{E} Journal}  \bvol{58},
  \pg{1381--1396}.

\bibitem[Schiller \& Naumann(1935)]{schiller1935drag}
{\sc \au{Schiller, L.} \& \au{Naumann, A.}} \yr{1935}  \at{A drag coefficient
  correlation}.  \jt{Zeitschrift des Vereins Deutscher Ingenieure}  \bvol{77},
  \pg{318--320}.

\bibitem[Shaffer {\em et~al.\/}(2013)Shaffer, Gopalan, Breault, Cocco, Karri,
  Hays \& Knowlton]{shaffer2013high}
{\sc \au{Shaffer, F.}, \au{Gopalan, B.}, \au{Breault, R.~W.}, \au{Cocco, R.},
  \au{Karri, S.~B.}, \au{Hays, R.} \& \au{Knowlton, T.}} \yr{2013}  \at{High
  speed imaging of particle flow fields in {CFB} risers}.  \jt{Powder
  Technology}  \bvol{242},  \pg{86--99}.

\bibitem[Sinclair \& Jackson(1989)]{sinclair1989gas}
{\sc \au{Sinclair, J.~L.} \& \au{Jackson, R.}} \yr{1989}  \at{Gas--particle
  flow in a vertical pipe with particle-particle interactions}.  \jt{AIChE
  Journal}  \bvol{35}~(9),  \pg{1473--1486}.

\bibitem[Spencer \& Rivlin(1958)]{Spencer1958}
{\sc \au{Spencer, A. J.~M.} \& \au{Rivlin, R.~S.}} \yr{1958}  \at{The theory of
  matrix polynomials and its application to the mechanics of isotropic
  continua}.  \jt{Archive for Rational Mechanics and Analysis}  \bvol{2},
  \pg{309--336}.

\bibitem[Speziale {\em et~al.\/}(1991)Speziale, Sarkar \& B.]{ML_1991Speziale}
{\sc \au{Speziale, C.~G.}, \au{Sarkar, S.} \& \au{B., Gatski~T.}} \yr{1991}
  \at{Modelling the pressure-strain correlation of turbulence: an invariant
  dynamical systems approach}.  \jt{Journal of Fluid Mechanics}  \bvol{227},
  \pg{245--272}.

\bibitem[Sun \& Zhu(2019)]{Sun2019}
{\sc \au{Sun, Z.} \& \au{Zhu, J.}} \yr{2019}  \at{A consolidated flow regime
  map of upward gas fluidization}.  \jt{AIChE Journal}  \bvol{65},  \pg{1--15}.

\bibitem[Sundaram \& Collins(1994)]{sundaram1994spectrum}
{\sc \au{Sundaram, S.} \& \au{Collins, L.~R.}} \yr{1994}  \at{Spectrum of
  density fluctuations in a particle-fluid system-{I}. {M}onodisperse spheres}.
   \jt{International Journal of Multiphase Flow}  \bvol{20}~(6),
  \pg{1021--1037}.

\bibitem[Tenneti \& Subramaniam(2011)]{Tenneti2011}
{\sc \au{Tenneti, S.} \& \au{Subramaniam, S.}} \yr{2011}  \at{Drag law for
  monodisperse gas-solid systms using particle-resolved direct numerical
  simulation of flow past fixed assemblies of spheres}.  \jt{International
  Journal of Multiphase Flow}  \bvol{37}~(9),  \pg{1072--1092}.

\bibitem[Tracey {\em et~al.\/}(2015)Tracey, Duraisamy \& Alonso]{ML_2015Tracey}
{\sc \au{Tracey, B.}, \au{Duraisamy, K.} \& \au{Alonso, J.~J.}} \yr{2015}
  \at{A machine learning strategy to assist turbulence model development}.
  \jt{AIAA Paper}  \bvol{1287}.

\bibitem[Zeng \& Zhou(2006)]{Zeng2006}
{\sc \au{Zeng, Zh.~X.} \& \au{Zhou, L.~X.}} \yr{2006}  \at{A two-scale
  second--order moment particle turbulence model and simulation of dense
  gas--particle flows in a riser}.  \jt{Powder Technology}  \bvol{162},
  \pg{27--32}.

\end{thebibliography}

\end{document}